\documentclass[twocolumn]{aastex62}

\def\Hi{H\,{\sc i}}
\def\Civ{C\,{\sc iv}}
\def\Mgii{Mg\,{\sc ii}}

\def\Cii{C\,{\sc ii}}
\def\xray{X-ray}
\def\cxo{{\it Chandra}}
\def\xmm{{\it XMM-Newton}}
 
\shorttitle{{\em Chandra} Observations of Reionization-Era Quasars}
\shortauthors{Wang et al.}
\begin{document}

\title{\Large \bf Revealing the Accretion Physics of Supermassive Black Holes at Redshift $z\sim7$ with {\em Chandra} and Infrared Observations}

\correspondingauthor{Feige Wang}
\email{feigewang@email.arizona.edu}

\author[0000-0002-7633-431X]{Feige Wang}
\altaffiliation{Hubble Fellow}
\affil{Steward Observatory, University of Arizona, 933 North Cherry Avenue, Tucson, AZ 85721, USA}

\author[0000-0003-3310-0131]{Xiaohui Fan}
\affil{Steward Observatory, University of Arizona, 933 North Cherry Avenue, Tucson, AZ 85721, USA}

\author[0000-0001-5287-4242]{Jinyi Yang}
\altaffiliation{Strittmatter Fellow}
\affil{Steward Observatory, University of Arizona, 933 North Cherry Avenue, Tucson, AZ 85721, USA}

\author[0000-0002-5941-5214]{Chiara Mazzucchelli}
\affil{European Southern Observatory, Alonso de C\'ordova 3107, Casilla 19001, Vitacura, Santiago 19, Chile}

\author[0000-0002-7350-6913]{Xue-Bing Wu}
\affil{Kavli Institute for Astronomy and Astrophysics, Peking University, Beijing 100871, China}
\affil{Department of Astronomy, School of Physics, Peking University, Beijing 100871, China}

\author[0000-0001-6239-3821]{Jiang-Tao Li}
\affil{Department of Astronomy, University of Michigan, 311 West Hall, 1085 S. University Ave, Ann Arbor, MI, 48109-1107, USA}

\author[0000-0002-2931-7824]{Eduardo Ba\~nados}
\affil{The Observatories of the Carnegie Institution for Science, 813 Santa Barbara Street, Pasadena, California 91101, USA}

\author[0000-0002-6822-2254]{Emanuele Paolo Farina}
\affiliation{Max Planck Institut f\"ur Astrophysik, Karl--Schwarzschild--Stra{\ss}e 1, D-85748, Garching bei M\"unchen, Germany}

\author[0000-0002-2579-4789]{Riccardo Nanni}
\affil{Department of Physics, University of California, Santa Barbara, CA 93106-9530, USA}

\author[0000-0001-9312-4640]{Yanli Ai}
\affil{College of Engineering Physics, Shenzhen Technology University, Shenzhen, 518118, China}

\author[0000-0002-1620-0897]{Fuyan Bian}
\affil{European Southern Observatory, Alonso de C\'ordova 3107, Casilla 19001, Vitacura, Santiago 19, Chile}

\author[0000-0003-0821-3644]{Frederick B. Davies}
\affil{Lawrence Berkeley National Laboratory, CA 94720-8139, USA}

\author[0000-0002-2662-8803]{Roberto Decarli}
\affil{INAF--Osservatorio di Astrofisica e Scienza dello Spazio, via Gobetti 93/3, I-40129, Bologna, Italy}

\author[0000-0002-7054-4332]{Joseph F. Hennawi}
\affil{Department of Physics, University of California, Santa Barbara, CA 93106-9530, USA}
\affil{Max Planck Institut f\"ur Astronomie, K\"onigstuhl 17, D-69117, Heidelberg, Germany}

\author[0000-0002-4544-8242]{Jan-Torge Schindler}
\affil{Max Planck Institut f\"ur Astronomie, K\"onigstuhl 17, D-69117, Heidelberg, Germany}

\author[0000-0001-9024-8322]{Bram Venemans}
\affil{Max Planck Institut f\"ur Astronomie, K\"onigstuhl 17, D-69117, Heidelberg, Germany}

\author[0000-0003-4793-7880]{Fabian Walter}
\affil{Max Planck Institut f\"ur Astronomie, K\"onigstuhl 17, D-69117, Heidelberg, Germany}

\begin{abstract}
X-ray emission from quasars has been detected up to redshift $z=7.5$, although only limited to a few objects at $z>6.5$. In this work, we present new {\it Chandra} observations of five $z>6.5$ quasars. By combining with archival {\it Chandra} observations of six additional $z>6.5$ quasars, we perform a systematic analysis on the X-ray properties of these earliest accreting supermassive black holes (SMBHs). We measure the black hole masses, bolometric luminosities ($L_{\rm bol}$), Eddington ratios ($\lambda_{\rm Edd}$), emission line properties, and infrared luminosities ($L_{\rm IR}$) of these quasars using infrared and sub-millimeter observations. Correlation analysis indicates that the X-ray bolometric correction (the factor that converts from \xray\ luminosity to bolometric luminosity) decreases with increasing $L_{\rm bol}$, and that the UV/optical-to-\xray\ ratio, $\alpha_{\rm ox}$, strongly correlates with $L_{\rm 2500\AA}$, and moderately correlates with $\lambda_{\rm Edd}$ and blueshift of \Civ\ emission lines. These correlations are consistent with those found in lower-$z$ quasars, indicating 
quasar  accretion physics does not evolve with redshift. We also find that  $L_{\rm IR}$ does not correlate with  $L_{\rm 2-10 keV}$ in these luminous distant quasars, suggesting that the ratio of the SMBH growth rate and their host galaxy growth rate in these early luminous quasars are different from those of local galaxies. A joint spectral analysis of the X-ray detected $z>6.5$ quasars yields an average X-ray photon index of $ \Gamma=2.32^{+0.31}_{-0.30}$, steeper than that of low-$z$ quasars. By comparing it with the $\Gamma-\lambda_{\rm Edd}$ relation, we conclude that the steepening of $\Gamma$ for quasars at $z>6.5$ is mainly driven by their higher Eddington ratios. 
\end{abstract}

\section{Introduction} \label{sec_intro}
Quasars, the most luminous type of active galactic nuclei (AGN), are believed to be powered by accreting supermassive black holes (SMBHs). The continuum and line emission from luminous quasars, over a large wavelength range, from optical to X-ray, can be characterized by several major components: the optical-to-ultraviolet (UV) continuum emission which is explained by a standard accretion disk extending down to the innermost stable circular orbit \citep[ISCO; e.g.][]{Shields78}, a soft \xray\ excess whose origin is still debated \citep[e.g.][]{Arnaud85}, \xray\ emission with a power-law spectrum produced by inverse Compton scattering of photons from the accretion disk of relativistic electrons in the hot corona 
\citep[e.g.][]{Svensson94}, and the broad emission lines 
emitted from the so-called broad line region \citep[BLR; e.g.][]{Antonucci93}.  Thus the optical/UV to \xray\ emission of quasars provide crucial information about the BH mass, the structure and physics of the accretion flow around the central SMBHs.

At present more than 200 quasars have been discovered at redshift $z>6$ \citep[e.g.][]{Fan01,Wu15,Jiang16,Banados16,Wang17,Reed17,Yang17}; about 50 quasars have been discovered at $z>6.5$ \citep[e.g.][]{Venemans13,Wang19b,Mazzucchelli17,Yang19} and seven at  $z>7$ \citep{Mortlock11, Banados18a, Wang18, Matsuoka19, Yang19, Yang20}. 
Extensive optical to near-infrared (NIR) spectroscopic observations of these quasars indicate that billion solar mass SMBHs are already in place when the universe is only $\sim$700 Myrs old \citep{Yang20}. The growth of these early SMBHs is limited by the available accretion time. At $z\sim7$, only $\sim14$ $e$-folding times elapsed since the first luminous object formed in the Universe \citep[i.e. $z\sim30$,][]{Tegmark97}, corresponding to a factor of $\sim10^6$ increase in mass, placing the most stringent constraints on the SMBH formation and growth mechanisms \citep[e.g.][]{Banados18a, Yang20}. In order to explain the existence of these SMBHs, many theoretical models have been proposed \citep[see][and references therein]{Latif16,Inayoshi20} by invoking either super-Eddington accretion process \citep[e.g.,][]{Volonteri15} and/or a massive seed BH \citep[e.g.,][]{Omukai08, Volonteri08, Wise19}. 

The \xray\ emission from quasars carries crucial information about the accretion physics and AGN feedback \citep[e.g.][]{Fabian14,Parker17}. However, \xray\ observations are only available for a very limited sample at high redshift. To date, $\sim$30 $z\sim$6 quasars \citep[e.g.,][]{Brandt01,Shemmer06,Ai16,Ai17,Nanni17,Nanni18,Vito19,Connor19} and six $z>6.5$ quasars \citep{Page14,Moretti14,Banados18b,Vito19,Pons20,Connor20} have been detected in \xray\ with \cxo\ and \xmm. Two key findings have been established based on these limited \xray\ observations. 

\begin{deluxetable*}{ccccccccccccc}
\setlength{\tabcolsep}{3pt}
\tablecaption{Basic properties and observations for $z>6.5$ quasars\label{tab_basic}}
\tabletypesize{\scriptsize}
\tablehead{\colhead{Name} & \colhead{Ra} & \colhead{Dec} & \colhead{$z$} & \colhead{$J_{\rm AB}$} & 
\colhead{obs. date} & \colhead{ObsID} & \colhead{Mode} & \colhead{${t_{\rm exp, X}}$} & \colhead{NIR Inst.} & \colhead{${t_{\rm exp, NIR}}$} & \colhead{$N_{\rm H}$} & \colhead{Ref. (disc./$z$)} }
\startdata
                         &                     &                         &             &                            & yyyy-mm-dd &           &              & [ks] &             &    [ks]    & [$10^{20}$ cm$^{-2}$]  & \\
\hline
J2348$-$3054 & 23:48:33.34 & $-$30:54:10.0 & 6.9018 & 21.11$\pm$0.11 & 2018-09-04 & 20414 & VFAINT & 42.50 & X-Shooter & 9.2 & 1.30 & V13/V16\\
J1048$-$0109 & 10:48:19.09 & $-$01:09:40.2 & 6.6759 & 20.61$\pm$0.17 & 2019-01-28 & 20415 & VFAINT & 34.76 & X-Shooter & 4.8 & 3.60 & W17/D18\\
J0024$+$3913 & 00:24:29.77 & $+$39:13:19.0 & 6.6210 & 20.70$\pm$0.15 & 2018-05-21 & 20416 & VFAINT & 19.70 & GNIRS & 13.8 & 6.76 & T17/M17\\
J2132$+$1217 & 21:32:33.19 & $+$12:17:55.3 & 6.5850 & 19.55$\pm$0.11 & 2018-08-20 & 20417 & VFAINT & 17.82 & X-Shooter & 8.4 & 6.42 &M17/D18\\
J0224$-$4711 & 02:24:26.54 & $-$47:11:29.4 & 6.5223 & 19.73$\pm$0.06 & 2018-03-05 & 20418 & VFAINT & 17.72 & X-Shooter & 4.8 & 1.66 & R17/W20\\
\hline
J1342$+$0928 & 13:42:08.11 & $+$09:28:38.6 & 7.5413 & 20.36$\pm$0.10 & 2017-12-15 & 20124 & VFAINT & 24.73 & GNIRS & 32.4 & 2.04 &B18/V17\\
                         &                     &                         &              &                            & 2017-12-17 & 20887 & VFAINT & 20.38 &             &             &       & \\
J1120$+$0641 & 11:20:01.48 & $+$06:41:24.3 & 7.0842 & 20.30$\pm$0.15 & 2011-02-04 & 13203 & FAINT & 15.84 & GNIRS & 4.8 & 5.07 &M11/D18\\
J2232$+$2930 & 22:32:55.15 & $+$29:30:32.0 & 6.6580  & 20.28$\pm$0.14 & 2018-01-30 & 20395 & VFAINT & 54.21 & GNIRS & 4.8 & 6.71 &V15/D18\\
J0305$-$3150 & 03:05:16.92 & $-$31:50:56.0 & 6.6145  & 20.70$\pm$0.09 & 2018-05-11 & 20394 & VFAINT & 49.88 & X-Shooter & 16.8 & 1.42 &V13/V16\\
J0226$+$0302 & 02:26:01.87 & $+$03:02:59.3 & 6.5412 & 19.43$\pm$0.10 & 2018-10-09 & 20390 & VFAINT & 25.90 & X-Shooter & 4.8 & 3.04 & V15/B15\\
J1110$-$1329 & 11:10:33.96 & $-$13:29:45.6 & 6.5148 & 21.16$\pm$0.09 & 2018-02-20 & 20397 & VFAINT & 59.33 & FIRE & 12.0 & 5.31 & V15/D18
\enddata
\tablecomments{The first section includes five quasars with new \cxo\ observations, while the second section represents six quasars with archival \xray\ observations. All redshift comes from the fitting of [\Cii] emission line. The sources are sorted by decreasing redshift. \\
{\bf References: } B15: \cite{Banados15}; B18: \cite{Banados18b}; D18: \cite{Decarli18}; M17: \cite{Mazzucchelli17}; R17: \cite{Reed17}; T17: \cite{Tang17}; V13: \cite{Venemans13}; V16: \cite{Venemans16}
W20: The [\Cii] redshift of this object is obtained from ALMA Cycle 6 observations (2018.1.01188.S, PI: Wang) (Wang et al. {\it in preparation}).}
\end{deluxetable*}

First, there is a tight correlation between the optical/UV-\xray\ luminosity ratio ($\alpha_{\rm ox}$) and the UV luminosity (i.e. $L_{\rm2500\AA}$), and it does not evolve from low redshift up to $z\sim6$ \citep[e.g.][]{Just07,Lusso16,Nanni17}. Recent investigations of several $z>6.5$ quasars \citep{Moretti14,Page14,Banados18a,Vito19} suggest that this relation might still hold in the epoch of reionization. Since $\alpha_{\rm ox}$ measures the relative importance of the hot corona versus the accretion disk, the steeper $\alpha_{\rm ox}$ in higher luminosity quasars indicates the dominance of the disk emission with respect to the hot electron corona emission in luminous quasars \citep[see][for a review]{Brandt15}.

The other key finding is that there is a moderate positive correlation between the photon index, $\Gamma$, of the hard \xray\ spectrum ($N(E) \propto E^{-\Gamma}$) and the Eddington ratio ($\lambda_{\rm Edd}=L/L_{\rm Edd}$)  established from a sizable sample of sources up to $z\sim3$, with larger $\Gamma$ corresponding to higher $\lambda_{\rm Edd}$ \citep[e.g.][but see \cite{Trakhtenbrot17}]{Shemmer08,Brightman13}. A high accretion rate is expected to increase the disk temperature and thus the level of disk emission, resulting in the increase of Compton cooling of the corona \citep[e.g.,][]{Maraschi97}, and producing a steep (large $\Gamma$) \xray\ spectrum. 
However, the relation between $\Gamma$ and $\lambda_{\rm Edd}$ is far from 
well established for the most distant quasars. Measuring $\Gamma$ is extremely difficult at high redshift because of the limited photon statistics. To date, 
only for four quasars (three at $z\sim6$ quasars and one at $z>7$) have more than 100 \xray\  photons been detected, which is required to place reasonable constraints
on $\Gamma$ for individual quasars \citep{Page14,Moretti14,Ai17,Nanni17,Nanni18}. 
Alternatively, stacking studies of quasars to study the average $\Gamma$ at differents redshifts indicates that the average $\Gamma$ does not evolve from $z\sim0$ to $z\sim6$ \citep{Just07,Vignali05,Shemmer06,Nanni17}. However, the more recent work by \cite{Vito19} indicates that the average $\Gamma$ of three $z>6.5$ quasars is slightly steeper than but still consistent with those of typical quasars at $z<6$.

In this paper, we report new \cxo\ observations of five quasars at $z>6.5$, significantly increasing the number of \xray\ observed quasars at these redshifts. Together with archival \cxo\ observations of six additional $z>6.5$ quasars, we perform joint spectral fitting of all \xray\ detected $z>6.5$ quasars with a mean quasar redshift of $z=6.822$. We also analyze the NIR spectra for these quasars and investigate the relations between quasar rest-frame UV and \xray\ properties. 
In Section \ref{sec_obs} we describe the \xray\ and NIR observations and data reduction. 
We present the \xray\ fluxes, luminosities, $\alpha_{\rm ox}$ measurements from \cxo\ observations, the black hole masses, bolometric luminosities, Eddington ratios and line properties measurements from NIR spectral fitting, and the infrared luminosities measured from sub-millimeter observations in Section \ref{sec_measure}.
The correlation between \xray\ and other properties of individual quasars are investigated in Section \ref{sec_correlation}. 
We present the stacked \xray\ spectrum, joint spectral fitting, and the mean properties of these $z>6.5$ quasars in Section \ref{sec_stack}. 
Finally, we conclude and summarize our findings in Section \ref{sec_summary}. 
Throughout the paper, we adopt a flat cosmological model with $H_0=68.5~{\rm km~s^{-1}~Mpc^{-1}}$  \citep{Betoule14}, $\rm \Omega_M=0.3$, and $\Omega_\Lambda=0.7$. All the uncertainties of our measurements reported in this work are at 1-$\sigma$ confidence level, while upper limits are reported at the 95\% confidence level.

\section{Observations and Data Reduction} \label{sec_obs}

\begin{figure*}[tbh]
\centering
\includegraphics[width=0.9\textwidth]{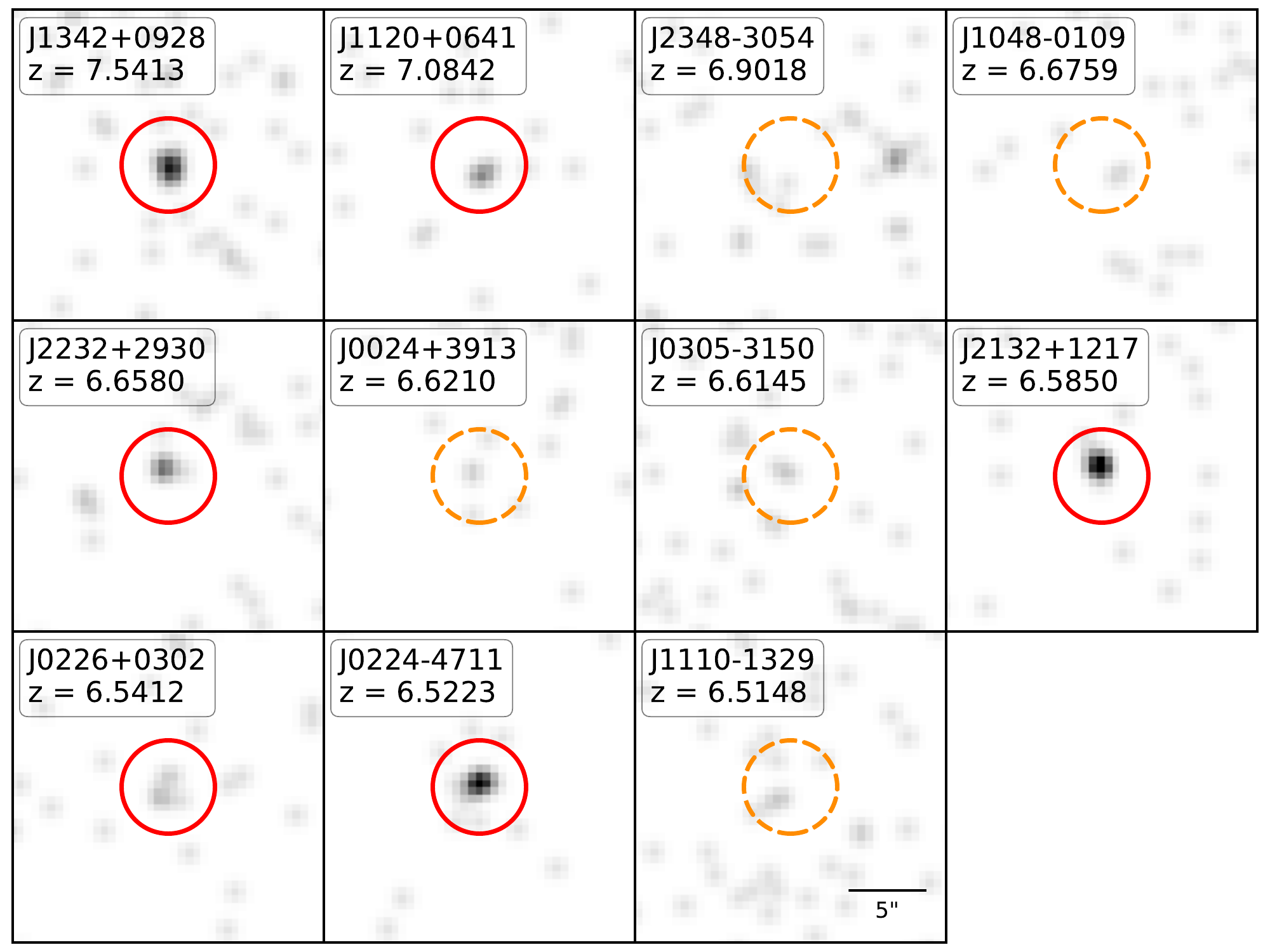}
\caption{Full-band (0.5--7 keV) \cxo\ cutouts of the eleven $z>6.5$ quasars. The images are centered at the optical positions listed in Table \ref{tab_basic}. The images have been smoothed with a 1 pixel Gaussian kernel. Red solid circles highlight targets detected by {\tt wavdetect}, while the orange dashed circles denote targets not detected by {\tt wavdetect}. All circles represent our extraction regions with a $3.0''$ radius. All cutouts are in the size of $20''\times20''$.
\label{fig_cxo}}
\end{figure*}

\subsection{{\it Chandra} \xray\ Observations}
We obtained {\it Chandra} observations of five quasars at $z>6.5$ using the Advanced CCD imaging spectrometer \citep[ACIS-S,][]{Garmire03} instrument in Cycle 19 (proposal number: 19700283, PI. Fan). 
The five quasars observed were 
J002429.77+391319.0 \citep[hereafter J0024+3913,][]{Tang17} at $z=6.6210$,
J022426.54--471129.4 \citep[hereafter J0224--4711,][]{Reed17}  at $z=6.5223$, 
J104819.09--010940.21 \citep[hereafter J1048--0109,][]{Wang17}  at $z=6.6759$,
J213233.19+121755.3 \citep[hereafter J2132+1217,][]{Mazzucchelli17}  at $z=6.5850$, and J234833.34--305410.0 \citep[hereafter J2348--3054,][]{Venemans13}  at $z=6.9018$.
These targets were positioned on the ACIS-S3 chip with the Very Faint telemetry format and the Timed Exposure mode. The observation log and the basic properties (i.e. redshift and brightness) of these quasars are listed in Table \ref{tab_basic}.

In order to increase the sample size of our analysis we also include the six other $z>6.5$ quasars that were observed by \cxo\ and archived as of 2020 April. Specifically, J112001.48+064124.3 \citep[hereafter J1120+0641,][]{Mortlock11}  at $z=7.09$ was observed in Cycle 12 \citep{Page14}, J134208.10 +092838.6 \citep[hereafter J1342+0928,][]{Banados18b} at $z=7.54$ was observed in Cycle 18 \citep{Banados18a}, and four other quasars were observed in Cycle 19 \citep{Vito19}. 
The observation log and properties of these quasars are also listed in Table \ref{tab_basic}. The Galactic \Hi\ column density at each quasar position calculated from \cite{Kalberla05} is also listed in Table \ref{tab_basic}. Similar with our new observations, these $z>6.5$ quasars were positioned on the ACIS-S3 chip with the Timed Exposure mode. J1120+0641 was observed with the Faint telemetry format and all other quasars were observed with the Very Faint mode.

The data were reprocessed with the {\tt chandra\_repro} script in the standard Chandra's data analysis system: CIAO \citep[][]{Fruscione06} version 4.12 and CALDB version 4.9.0. In the analyses, only grade 0, 2, 3, 4, and 6 events were used. In the process, we set the option {\tt check\_vf\_pha = yes} in the case of observations taken in very faint mode. The exposure maps and the PSF maps were created with the {\tt fluximage} script and the {\tt mkpsfmap} script, respectively. 
Considering the increasingly uncertain quantum efficiency of ACIS at lower energies and the steeply increasing background at higher energies, we only used the \xray\ counts at observed frame energies of 0.5--7keV, following \cite{Nanni17}. In order to detect sources we first performed source detections using {\tt wavdetect} \citep{Freeman02} with a false-positive probability threshold of $10^{-5}$. Six quasars were detected by {\tt wavdetect}: J1342+0928, J1120+0641, J2232+2930, J2132+1217, J0226+0302, and J0224--4711, with the net counts of $14.2^{+5.0}_{-3.8}$, $5.8^{+3.6}_{-2.4}$, $7.1^{+3.9}_{-2.8}$, $15.6^{+5.1}_{-4.0}$, $5.5^{+3.6}_{-2.4}$, and $18.3^{+5.4}_{-4.3}$, respectively. The uncertainties are estimated according to the approximation of \cite{Gehrels86}.

\begin{deluxetable*}{crrrrrrrrrr}                                                                      
\tablecaption{ \xray\ photometry and quasar \xray\ properties.\label{tab_cxo}}
\tablehead{
\colhead{Name} &
  \multicolumn{3}{c}{Net Counts}&
\colhead{HR} &
  \multicolumn{3}{c}{Flux \tablenotemark{b}} &
\colhead{$L_{\rm 2-10keV}$} & \colhead{$\rm\alpha_{ox}$}  & \\
\cline{2-4} \cline{6-8}
\colhead{} &
\colhead{0.5--7.0 keV} &
\colhead{0.5--2.0 keV} &
\colhead{2.0--7.0 keV} &
\colhead{} & 
\colhead{0.5--7.0 keV} &
\colhead{0.5--2.0 keV} &
\colhead{2.0--7.0 keV} &
\colhead{$\rm 10^{44}~erg~s^{-1}$} & 
\colhead{} 
}
\startdata
J1342$+$0928 & $14.0^{+5.1}_{-4.0}$ & $10.2^{+4.4}_{-3.3}$ & $3.8^{+3.4}_{-2.2}$ & $-0.46^{+0.23}_{-0.29}$   & $2.97^{+1.08}_{-0.85}$ & $1.56^{+0.57}_{-0.45}$ & $1.41^{+0.51}_{-0.40}$ & $12.80^{+4.68}_{-3.69}$ & $-1.61^{+0.05}_{-0.06}$& \\  
J1120$+$0641 & $5.3^{+3.6}_{-2.4}$ & $3.7^{+3.2}_{-1.9}$ & $1.6^{+2.6}_{-1.3}$ & $-0.37^{+0.33}_{-0.47}$   & $2.29^{+1.56}_{-1.04}$ & $1.20^{+0.82}_{-0.54}$ & $1.09^{+0.74}_{-0.49}$ & $8.52^{+5.82}_{-3.83}$& $-1.66^{+0.09}_{-0.10}$& \\  
J2348$-$3054 & $<8.6$\tablenotemark{a} & $<3.0$ & $<9.7$ & -- & $<2.26$ & $<1.19$ & $<1.07$ & $<7.96$ & $<-1.56$ \\  
J1048$-$0109 & $<4.7$ & $<4.1$ & $<4.0$ & -- & $<1.54$ & $<0.81$ & $<0.73$ & $<5.02$ & $<-1.72$ \\  
J2232$+$2930 & $6.6^{+4.1}_{-2.9}$ & $6.3^{+3.8}_{-2.6}$ & $0.4^{+2.6}_{-1.3}$ & $-0.73^{+0.07}_{-0.27}$  & $1.45^{+0.90}_{-0.64}$ & $0.76^{+0.47}_{-0.33}$ & $0.69^{+0.43}_{-0.30}$ & $4.68^{+2.89}_{-2.03}$& $-1.73^{+0.08}_{-0.09}$& \\  
J0024$+$3913 & $<8.2$ & $<5.9$ & $<5.6$ & -- & $<4.93$ & $<2.59$ & $<2.34$ &  $<15.74$ & $<-1.47$ & \\  
J0305$-$3150 & $<5.8$ & $<4.2$ & $<5.1$ & -- & $<1.31$ & $<0.69$ & $<0.62$ &  $<4.18$ & $<-1.73$ &\\  
J2132$+$1217 & $15.0^{+5.1}_{-4.0}$ & $7.5^{+3.9}_{-2.8}$ & $7.5^{+3.9}_{-2.8}$ &  $0.00^{+0.27}_{-0.25}$  & $9.93^{+3.38}_{-2.65}$ & $5.22^{+1.77}_{-1.39}$ & $4.71^{+1.60}_{-1.26}$ & $31.32^{+10.62}_{-8.34}$& $-1.50^{+0.05}_{-0.05}$&  \\  
J0226$+$0302 & $5.0^{+3.6}_{-2.4}$ & $3.7^{+3.2}_{-1.9}$ & $1.2^{+2.6}_{-1.3}$ &  $-0.49^{+0.16}_{-0.51}$ & $2.18^{+1.57}_{-1.05}$ & $1.14^{+0.82}_{-0.55}$ & $1.03^{+0.74}_{-0.49}$ & $6.73^{+4.84}_{-3.25}$& $-1.81^{+0.09}_{-0.11}$& \\  
J0224$-$4711 & $18.1^{+5.4}_{-4.3}$ & $15.7^{+5.1}_{-4.0}$ & $2.4^{+2.9}_{-1.6}$ & $-0.72^{+0.12}_{-0.21}$  & $11.44^{+3.41}_{-2.72}$ & $6.01^{+1.79}_{-1.43}$ & $5.43^{+1.62}_{-1.29}$ &$35.26^{+10.5}_{-8.39}$ & $-1.55^{+0.04}_{-0.05}$&  \\  
J1110$-$1329 & $<8.0$ & $<3.0$ & $<8.8$ & -- & $<1.58$ & $<0.83$ & $<0.75$ & $<4.86$ & $<-1.60$ &
\enddata
\tablenotetext{a}{For undetected objects, we report the upper limit corresponding to the 95\% confidence interval.}
\tablenotetext{b}{The Galactic absorption-corrected \xray\ flux in the observed band in units of $\rm 10^{-15}~erg~cm^{-2}~s^{-1}$.}
\tablecomments{The sources are sorted by decreasing redshift. The last column, $\rm\alpha_{ox}$, is measured using Equation (2) and the $L_{\rm 2500\AA}$} are listed in Table \ref{tab_nir}. The $\rm \Delta \alpha_{ox}$ used in Figure \ref{fig_delta_ox} can be derived by subtracting $\rm\alpha_{ox}$ from the $\alpha_{\rm ox}-L_{\rm 2500\AA}$ relation in \cite{Timlin20}.
\end{deluxetable*}

We extract the spectrum for each object within a $3\farcs0$ radius circular region centered at the optical position using the {\tt specextract} script. We choose a background annulus centered at the optical positions with an inner radius of $10\farcs0$ and an outer radius of $30\farcs0$. The net \xray\ counts detected in the soft band (0.5--2 keV), the hard band (2--7 keV), and the full band (0.5--7 keV) within the $3\farcs0$ radius circular region are reported in Table \ref{tab_cxo}. For undetected sources we report the 2$\sigma$ upper limits (corresponding to the 95\% confidence intervals) computed from the {\tt srcflux} script in CIAO. 
Table \ref{tab_cxo} also lists the hardness ratio ${\rm HR} = (H-S)/(H+S)$, where H and S are the net counts in the hard (2--7 keV) and soft (0.5--2.0 keV) bands, respectively. The HR for those \cxo\ detected quasars are estimated with the Bayesian method described by \cite{Park06}. The full band (0.5-7 keV) image stamps are shown in Figure \ref{fig_cxo}.

\begin{figure*}
\centering
\includegraphics[width=1.0\textwidth]{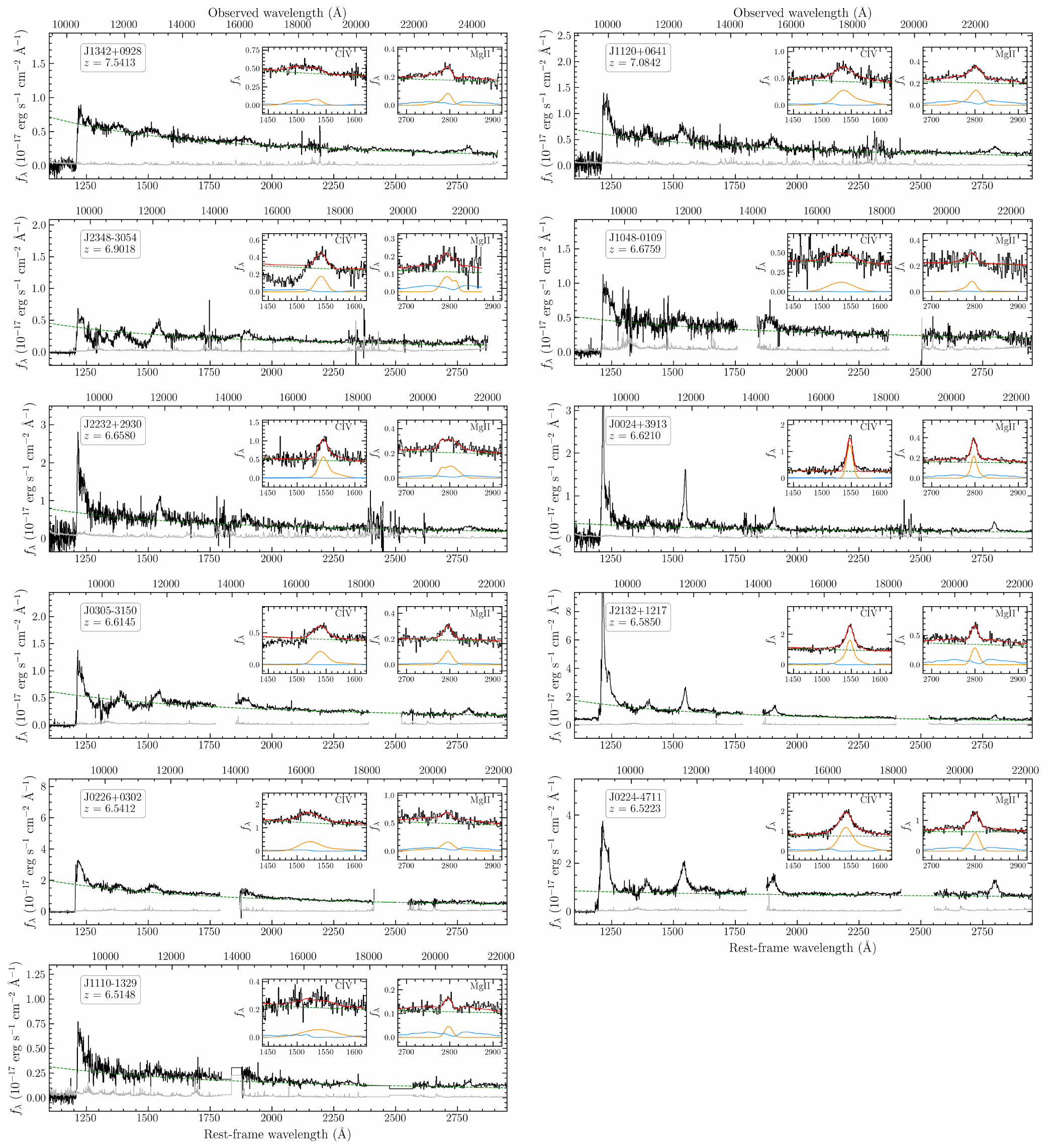}
\caption{Near-infrared spectra of the eleven $z>6.5$ quasars studied in this paper. The spectra are re-binned to 200 $\rm km~ s^{-1}$ pixels. The black and gray lines represent the Galactic extinction corrected spectra and one sigma error vectors. The green dashed lines denote best-fit UV power-law continuum. The left insert panels are the zoom-in of \Civ\ line fitting, while the right insert panels are the zoom-in of \Mgii\ line fitting. In the insert panels, red solid lines denote total fit, green dashed lines denote power-law continuum, blue lines denote iron template, and the orange lines denote emission line components. 
\label{fig_spec}}
\end{figure*}

\subsection{Near-Infrared Spectroscopy} \label{sec_nir}

We note that most of the quasars investigated here have BH mass estimates in the literature \citep[][]{Mortlock11, Derosa14, Mazzucchelli17, Banados18b, Tang19, Onoue20, Schindler20}.
However, these estimates were based on different fitting algorithms and different single-epoch virial scaling relations. To reduce the biases introduced
by different methods we perform our own  self-consistent  measurements of the masses and Eddington ratios of these SMBHs. We reduced and analyzed the archival NIR spectroscopic observations of these quasars. The quasar J1342+0928, J1120+0641, J0024+3913, and J2232+2930 were observed with Gemini/GNIRS \citep{gnirs1, gnirs2} using the Cross-dispersed mode. J1110--1329 was observed with Magellan/FIRE \citep{fire} using the Echelle mode. All the other quasars presented in this work were observed with VLT/X-Shooter \citep{xshooter}.

We reduced both GNIRS and X-Shooter spectra with the newly developed open source spectroscopic data reduction pipeline {\tt PyPeIt}\footnote{\url{https://github.com/pypeit/PypeIt}} \citep{pypeit1,pypeit2}. 
The wavelength solutions were derived from the night sky OH lines in the vacuum frame. The sky background was subtracted with the standard A--B mode and then a $b$-spline fitting procedure was performed to further clean up the sky line residuals following \cite{Bochanski09}. An optimal extraction \citep{Horne86} is then performed to generate 1D science spectra. We flux the extracted spectra with sensitivity functions derived from standard star observations. We then stacked the fluxed individual exposures and individual orders. The telluric corrections are performed by jointly fitting the atmospheric models derived from the Line-By-Line Radiative Transfer Model \citep[{\tt LBLRTM}\footnote{\url{http://rtweb.aer.com/lblrtm.html}};][]{Clough05} and a quasar model based on a Principal Component Analysis method \citep{Davies18} to the stacked quasar spectra. We then scaled the telluric corrected spectra to match the $J$-band photometry of each object by carrying out synthetic photometry on the spectrum for the purpose of absolute flux calibration. Finally, we corrected the Galactic extinction based on the dust map \citep{SFD98} and extinction law \citep{EXT89}. The fully calibrated NIR spectra of these quasars are shown in Figure \ref{fig_spec}.
The FIRE spectrum was reduced with the standard {\tt FIREHOSE} pipeline, which evolved from the MASE pipeline for optical echelle reduction \citep{Bochanski09}. We corrected for telluric absorption features by obtaining a spectrum of an A0V star at a comparable observing time.

\section{Measurements and Results} \label{sec_measure}
Since all quasars have less than 20 net counts in the full 0.5-7.0 keV band, we do not attempt spectral fitting for individual quasars. We measure the \xray\ flux by assuming a power-law spectrum with $\rm \Gamma=2.0$ \citep[typical of luminous quasars, e.g., ][]{Nanni17,Vito19}, accounting for the Galactic absorption \citep{Kalberla05}, and using the response matrices and ancillary files extracted at the position of each target. The rest-frame 2-10 keV luminosites were estimated by assuming $\rm \Gamma=2.0$ as listed in Table \ref{tab_cxo}. The measured \xray\ luminosity of these quasars spans more than an order of magnitude with  $L_{\rm 2-10keV}\lesssim 4-35\times10^{44}~ {\rm erg~s^{-1}}$. Note that the $L_{\rm 2-10keV}$ would be $\sim$20\% higher if we use $\rm \Gamma=2.3$, the average photon index of $z>6.5$ quasars derived from \S \ref{sec_stack}. Considering that most previous work has used $\rm \Gamma\sim2$ when measuring $L_{\rm 2-10keV}$ at high redshifts \citep[e.g.][]{Nanni17,Vito19}, we will only use the $L_{\rm 2-10keV}$ values derived by assuming $\rm \Gamma=2$ in what follows. 

To derive the rest-frame ultraviolet (UV) luminosities, black hole masses, and Eddington ratios for these quasars  we performed a global spectral fitting on the de-redshifted NIR spectra following \cite{Wang20}. Briefly, we first fit a pseudo-continuum model to the emission line (except for iron emission) free regions. The pseudo-continuum model includes three components, a power-law continuum ($f_\lambda \propto \lambda^{\alpha_\lambda} $), Balmer continuum \citep[e.g.][]{Derosa14}, and iron emission \citep{Vestergaard01,Tsuzuki06}. The iron template was constructed by composing the iron emission from \cite{Tsuzuki06} (2200\AA-3500\AA) and \cite{Vestergaard01} (1100\AA-2200\AA). The \Mgii\ and \Civ\ lines are then fitted with two Gaussian functions for each line after subtracting the pseudo-continuum model. We perform the whole fitting process iteratively and broaden the iron template by convolving it with a Gaussian kernel to match the line width of the \Mgii\ line. Following \cite{Wang20}, we use a Monte Carlo approach to estimate the spectral measurement uncertainties. We created 100 mock spectra by randomly adding Gaussian noise to each pixel with standard deviation equal to the spectral error at that pixel. Then we applied the exactly same fitting procedure to these mock spectra. The uncertainties of measured spectral properties are then estimated as the average of the 16\% and 84\% percentile deviation from the median. 

The derived power-law continuum slopes ($\alpha_\lambda$), continuum luminosities at rest-frame 2500 \AA, line widths, and redshifts are given in Table \ref{tab_nir}. 
The \Mgii\ and \Civ\ redshifts listed in Table \ref{tab_nir} were estimated based on the peak of the Gaussian fitting of each line. The redshifts based on the \Mgii\ line are in generally consistent with ($\rm<800~km~s^{-1}$) the [\Cii] redshifts listed in Table \ref{tab_basic}. The \Civ\ lines of all the quasars exhibit large blueshifts relative to both [\Cii] and \Mgii\ lines which we discuss in detail in \S \ref{subsec_ox}.
The bolometric luminosities are estimated by assuming a bolometric correction of  $L_{\rm bol}$=5.15$\times$ $\lambda L_{\rm3000 \text{\normalfont\AA}}$ \citep{Shen11}. The black hole masses, $M_{\rm BH}$,  are then estimated using the single virial estimator proposed by \cite{Vestergaard09}:
\begin{equation}
\small
\frac{M_{\rm BH}}{M_\odot} = 10^{6.86} \left[\frac{\lambda L_{\lambda}\rm{(3000~\AA)}}{\rm{10^{44}~ erg~s^{-1}}}\right]^{0.5}  \left[\frac{\rm{FWHM_{(Mg \ II)}}}{\rm{10^3~km~s^{-1}}}\right]^2
\end{equation}
The Eddington ratio of each quasar is then calculated as $\lambda_{\rm Edd}= L_{\rm bol}/L_{\rm Edd}$, where the $L_{\rm Edd}=1.26\times10^{38} ~M_{\rm BH}$ is the Eddington luminosity. The $M_{\rm BH}$ and $\lambda_{\rm Edd}$ are listed in Table \ref{tab_nir}. Note that the quoted uncertainties of $M_{\rm BH}$ and $\lambda_{\rm Edd}$ do not include the systematic uncertainties in the scaling relation, which is $0.55$ dex \citep{Vestergaard09}.

With the \xray\ and UV luminosity, we can then measure the optical-X-ray power-law slope, which is defined as  
\begin{equation}
\alpha_{\rm ox} = \frac{log(f_{\rm 2 \: keV}/f_{\rm 2500 \: \AA})}{log(\nu_{\rm 2 \: keV}/\nu_{\rm 2500 \: \AA})},
\end{equation}
where $f_{\rm 2 \: keV}$ and $f_{\rm 2500 \: \AA}$ are the flux densities at rest-frame 2 keV and 2500 \AA, respectively. The computed $\alpha_{\rm ox}$ values are listed in Table \ref{tab_cxo}. 

\begin{deluxetable*}{ccccccccccccc}[tbh]
\setlength{\tabcolsep}{1.5pt}
\tablecaption{Quasar properties derived from near-infrared and sub-millimeter observations. \label{tab_nir}}
\tabletypesize{\scriptsize}
\tablehead{\colhead{Name} & \colhead{$\rm M_{1450}$} & \colhead{$z_{\rm MgII}$} & \colhead{$z_{\rm CIV}$} & \colhead{$\rm FWHM_{MgII}$} & \colhead{$\rm FWHM_{CIV}$} & \colhead{$\lambda L_{\rm 2500}$} & \colhead{$L_{\rm bol}$} & \colhead{$\rm M_{BH}$} & \colhead{$\lambda_{\rm Edd}$} &  \colhead{$\rm S_{1mm}$} & \colhead{$L_{\rm IR}$} & \colhead{Ref. ($\rm S_{1mm}$})  }
\startdata
                        &                              &                               &                               & $\rm km~s^{-1}$ & $\rm km~s^{-1}$ & $\rm 10^{46}~erg~s^{-1}$ & $\rm 10^{47}~erg~s^{-1}$ & $\rm 10^{9}~M_\odot$ & & mJy &  $\rm 10^{46}~erg~s^{-1}$ \\
\hline
J1342$+$0928 & $-$26.65 & 7.531$\pm$0.004 & 7.361$\pm$0.025 & 2680$\pm$255 & 11776$\pm$1102 & 3.04$\pm$0.23 & 1.42$\pm$0.12 & 0.86$\pm$0.20 & 1.26$\pm$0.16 & 0.41$\pm$0.07 & 0.57$\pm$0.10 & V18\\
J1120$+$0641 & $-$26.45 & 7.095$\pm$0.002 & 7.027$\pm$0.005 & 3454$\pm$66 & 7554$\pm$690 & 2.80$\pm$0.24 & 1.35$\pm$0.11 & 1.40$\pm$0.10 & 0.74$\pm$0.06 & 0.53$\pm$0.04 & 0.73$\pm$0.06 & V18\\
J2348$-$3054 & $-$25.84 & 6.887$\pm$0.005 & 6.866$\pm$0.001 & 4385$\pm$786 & 4364$\pm$625 & 1.45$\pm$0.17 & 0.68$\pm$0.09 & 1.60$\pm$0.69 & 0.33$\pm$0.11 & 1.92$\pm$0.14 & 2.64$\pm$0.19 & V18\\
J1048$-$0109 & $-$26.03 & 6.661$\pm$0.005 & 6.603$\pm$0.015 & 2676$\pm$1240 & 10202$\pm$536 & 2.34$\pm$0.28 & 1.21$\pm$0.12 & 0.79$\pm$0.58 & 1.17$\pm$0.37 & 2.84$\pm$0.04 & 3.90$\pm$0.05 & D18\\
J2232$+$2930 & $-$26.34 & 6.666$\pm$0.007 & 6.642$\pm$0.001 & 5234$\pm$321 & 3938$\pm$215 & 2.35$\pm$0.40 & 1.11$\pm$0.19 & 2.91$\pm$0.57 & 0.29$\pm$0.02 & 0.97$\pm$0.22 & 1.33$\pm$0.30 & V18\\
J0024$+$3913 & $-$25.62 & 6.618$\pm$0.001 & 6.613$\pm$0.001 & 1711$\pm$139 & 2441$\pm$5 & 1.64$\pm$0.21 & 0.85$\pm$0.10 & 0.27$\pm$0.04 & 2.40$\pm$0.38 & 0.55$\pm$0.18 & 0.76$\pm$0.25 & V18\\
J0305$-$3150 & $-$26.11 & 6.608$\pm$0.002 & 6.576$\pm$0.003 & 2617$\pm$609 & 5624$\pm$300 & 2.08$\pm$0.13 & 1.01$\pm$0.07 & 0.70$\pm$0.38 & 1.12$\pm$0.44 & 3.29$\pm$0.10 & 4.52$\pm$0.14 & V18\\
J2132$+$1217 & $-$27.08 & 6.588$\pm$0.001 & 6.578$\pm$0.001 & 2146$\pm$263 & 3063$\pm$22 & 3.97$\pm$0.21 & 1.77$\pm$0.09 & 0.62$\pm$0.17 & 2.20$\pm$0.50 & 0.47$\pm$0.15 & 0.65$\pm$0.21 & V18\\
J0226$+$0302 & $-$27.26 & 6.532$\pm$0.017 & 6.427$\pm$0.003 & 3713$\pm$289 & 9346$\pm$1298 & 5.36$\pm$0.27 & 2.50$\pm$0.13 & 2.20$\pm$0.39 & 0.87$\pm$0.10 & 2.50$\pm$0.50 & 3.44$\pm$0.69 & V18\\
J0224$-$4711 & $-$26.67 & 6.527$\pm$0.001 & 6.486$\pm$0.001 & 2655$\pm$144 & 5760$\pm$96 & 5.85$\pm$0.35 & 3.36$\pm$0.20 & 1.30$\pm$0.18 & 1.98$\pm$0.15 & 1.96$\pm$0.07 & 2.70$\pm$0.10 & W20\\
J1110$-$1329 & $-$25.35 & 6.511$\pm$0.004 & 6.465$\pm$0.020 & 2267$\pm$352 & 13778$\pm$4155 & 1.10$\pm$0.14 & 0.55$\pm$0.06 & 0.38$\pm$0.14 & 1.10$\pm$0.24 & 0.87$\pm$0.05 & 1.20$\pm$0.07 & D18
\enddata
\tablecomments{The sources are sorted by decreasing redshift. \\
{\bf References: } D18: \cite{Decarli18}; V18: \cite{Venemans18}
W20: The 1mm continuum flux density of this object is obtained from ALMA Cycle 6 observations (2018.1.01188.S, PI: Wang) (Wang et al. {\it in preparation}).}
\end{deluxetable*}

In order to measure the infrared luminosities of these quasars we used the 1mm (in the observed frame) ALMA observations collected by \cite{Venemans18} and \cite{Decarli18}. In addition, we observed one quasar in our sample, J0224--4711, with ALMA in Cycle 6 (2018.1.01188.S, PI: Wang). In this paper, we only use the [\Cii] based redshift (Table \ref{tab_basic}) and the 1mm continuum (Table \ref{tab_nir}) measurements, while the detailed data reduction of our ALMA observations will be presented elsewhere (Wang et al. {\it in preparation}). 
Since Haro 11, a low metallicity dwarf galaxy, has been suggested as the best candidate analog for high-$z$ quasar host galaxies \citep[e.g.][]{Lyu16}, we estimate the 8--1000 $\mu$m infrared luminosities ($L_{\rm IR}$) of these quasar host galaxies by scaling the observed 1mm continuum to the Haro 11 spectral energy distribution (SED). 
The estimated $L_{\rm IR}$ are listed in Table \ref{tab_nir}. The star formation rate (SFR) can be calculated as 
\begin{equation}
{\rm SFR(M_\odot~yr^{-1})}=5.0\times10^{-44}~L_{\rm (IR, erg~s^{-1})}, 
\end{equation}
\citep{Lyu16}.
Note that the $L_{\rm IR}$ estimated using the Haro 11 template is usually about two times higher than that estimated from a modified blackbody with $T=47$ K and $\beta=1.6$ \citep{Beelen06}, because the modified blackbody misses flux in the mid-infrared. 

\section{Correlations Between X-ray Emission and Other Properties of Individual Quasars} \label{sec_correlation}
In this Section, we investigate the relationships between \xray\ emission and other properties of these high redshift quasars. Since our quasar sample is relatively small and only occupies the bright end (i.e. $L_{\rm bol}\gtrsim5\times10^{46}~{\rm erg~s^{-1}}$) of the quasar population at very high redshift, we consider a sample of $\sim2,000$ SDSS quasars at $1.7\le z \le 2.7$ that have \cxo\ observations \citep{Timlin20} to expand both sample size, luminosity range, and redshift range. We further restrict the redshift to be $z\ge2.0$ to ensure that we have the same rest-frame UV spectral coverage as the $z>6.5$ quasars studied here; this results in a sample of 1,175 objects. We then perform exactly the same spectral fitting method used in \S\, \ref{sec_measure} to compute the UV luminosities, $L_{\rm bol}$, BH masses, and Eddington ratios of these lower redshift quasars. 
We successfully fit the \Mgii\ emission lines in 897 of 1,175 objects, as some SDSS spectra have very low quality, some are strongly affected by residuals from OH sky lines, and some were obtained from the earlier SDSS spectrograph, which does not fully cover the wavelength range.

The $L_{\rm 2-10keV}$ of these SDSS quasars are adopted from \cite{Timlin20} and converted to the cosmological model used in this paper. The $\alpha_{\rm ox}$ are then calculated using $f_{\rm 2500 \: \AA}$ from our spectra fitting and $f_{\rm 2 \: keV}$ from \cite{Timlin20}. In order to determine the $L_{\rm IR}$ and SFR of SDSS quasars, we cross-matched the SDSS quasars from \cite{Timlin20} with the {\it Herschel}/SPIRE Point Source Catalogue (SPSC\footnote{https://doi.org/10.5270/esa-6gfkpzh}). To maximize the number of objects having both \xray\ and {\it Herschel} observations, we used the full sample of $\sim2,000$ quasars from \cite{Timlin20} for the matching. 
There are $\sim400$ quasars within the {\it Herschel}/SPIRE pointings but only 61 ($\sim$15\%) have been detected in at least one of the three bands (250$\mu$m, 350$\mu$m, and 500 $\mu$m). Thus, the 61 quasars only represent the far-infrared bright quasar population limited by the shallow {\it Herschel} observations. The $L_{\rm IR}$ of these SDSS quasars were then measured by fitting the SPIRE photometry to the Haro 11 SED, similar to the method used for $z>6.5$ quasars.
We also collected the 1mm observations for \xray\ detected $6<z<6.5$ quasars \citep{Vito19} from \cite{Venemans18} and \cite{Decarli18} and then measured the $L_{\rm IR}$ and SFR of these $6<z<6.5$ quasars using the same method for the $z>6.5$ quasars. 

\subsection{X-ray Bolometric Correction}\label{subsec_kbol}

\begin{figure}
\centering
\includegraphics[width=0.49\textwidth]{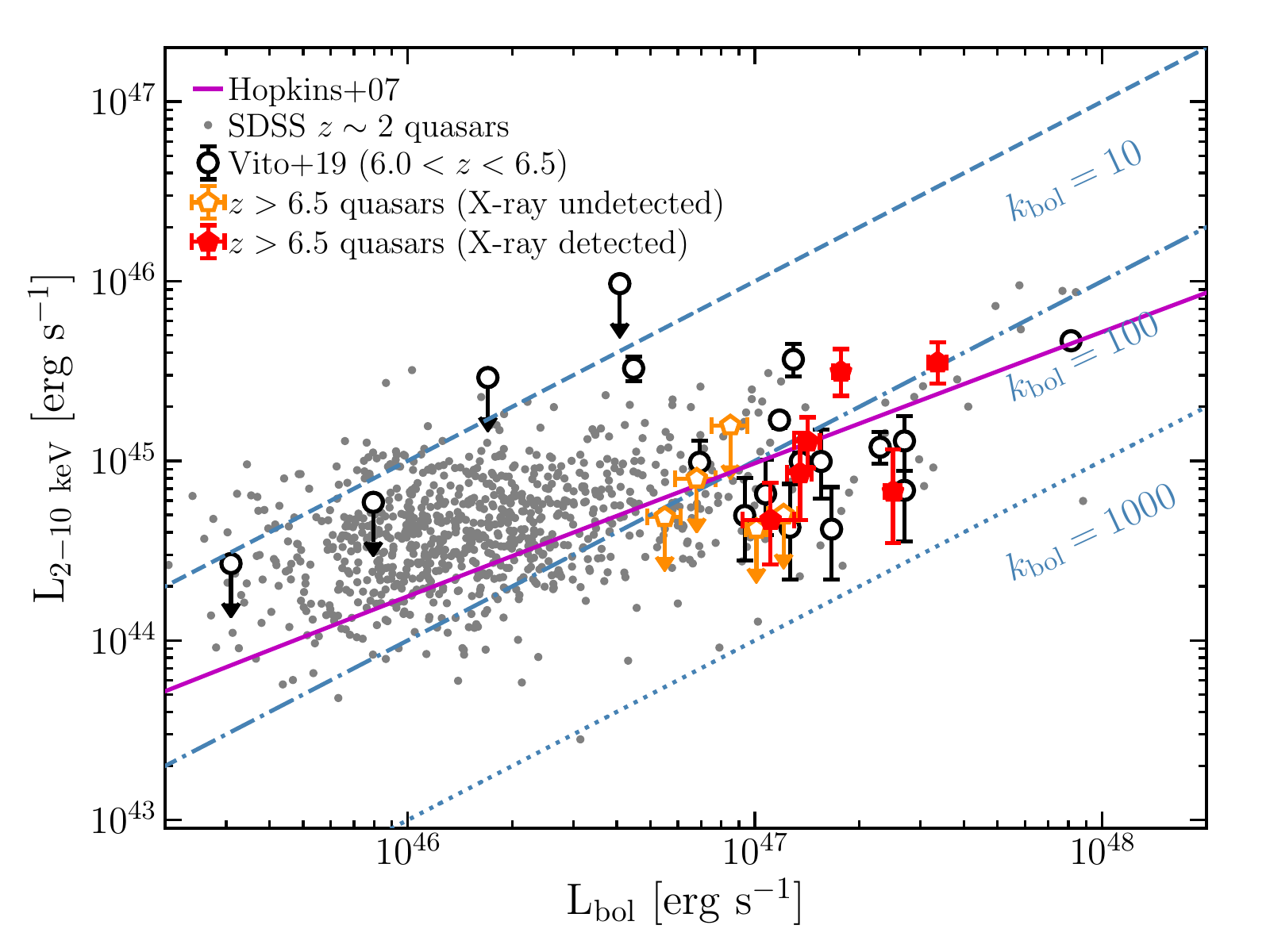}
\caption{\xray\ luminosity, $L_{\rm 2-10keV}$, versus bolometric luminosity, $L_{\rm bol}$, of $z>6.5$ quasars and SDSS $z\sim2$ quasars. The red pentagons represent \xray\ detected $z>6.5$ quasars, while the open orange pentagons denote \xray\ undetected $z>6.5$ quasars. The open circles denote quasars at $6.0<z<6.5$ from \cite{Vito19} and the small grey dots denote SDSS $z\sim2$ quasars from \cite{Timlin20}. The bolometric correction from $L_{\rm 2-10keV}$ of these $z>6.5$ quasars are $k_{\rm bol}\sim100$, consistent with the that in lower redshift quasars with similar luminosities \citep[e.g.][]{Hopkins07,Martocchia17}. 
\label{fig_lx}}
\end{figure}

Determining the relationship between \xray\ luminosity and bolometric luminosity is crucial in estimating the AGN bolometric luminosity function \citep[e.g.][]{Hopkins07} and the mass function of SMBH \citep[e.g.][]{Marconi04}. The relation between $L_{\rm 2-10keV}$ and $L_{\rm bol}$ has been well studied and an increasing bolometric correction $k_{\rm bol} = L_{\rm bol}/L_{\rm 2-10keV}$ with bolometric luminosity has been suggested \citep[e.g.][]{Marconi04, Hopkins07, Martocchia17}. 
In Figure \ref{fig_lx}, we show the relation between $L_{\rm 2-10keV}$ and $L_{\rm bol}$ of 11 $z>6.5$ quasars, 18 $6<z<6.5$ quasars from \cite{Vito19} as well as 897 SDSS $z\sim2$ quasars. 
The $k_{\rm bol}$ of most SDSS $z\sim2$ quasars are in the range of $10\lesssim k_{\rm bol} \lesssim100$, with the most luminous ones at $k_{\rm bol}\sim100$. 
The 11 $z>6.5$ quasars have a bolometric luminosity range of $\rm 0.5-3.4\times10^{47} \, erg \, s^{-1}$ and a \xray\ luminosity range of $\rm \lesssim0.5-3.5\times10^{45} \, erg \, s^{-1}$, suggesting $k_{\rm bol}\sim100$. The $k_{\rm bol}$ of these $z>6.5$ quasars is similar to that of the most luminous  SDSS $z\sim2$ quasars and $z\sim6$ quasars, in agreement with previous studies \citep[e.g.][]{Hopkins07,Vito19}, suggesting a redshift independent relationship between $k_{\rm bol}$ and $L_{\rm bol}$.  
 
 \subsection{Optical/UV to X-ray Flux Ratio, $\alpha_{\rm ox}$}\label{subsec_ox}

 \begin{figure}[tbh]
\centering
\includegraphics[width=0.48\textwidth]{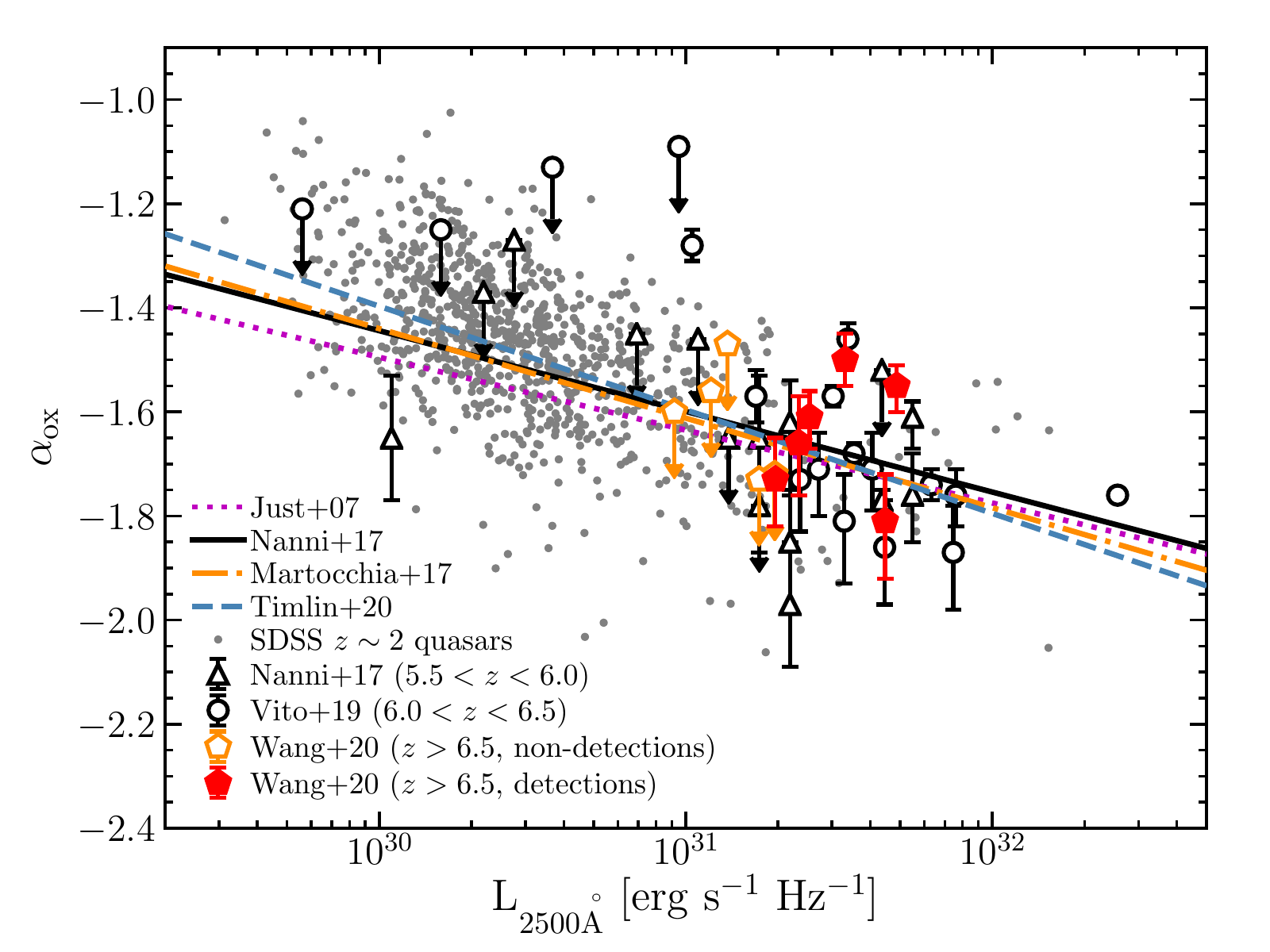}
\caption{The $\rm \alpha_{ox}$ versus $L_{\rm 2500\AA}$ plot. All symbols have the same meaning as Figure \ref{fig_lx}, except that we also include quasars at $5.5<z<6.0$ from \cite{Nanni17} as indicated by open triangles. All data points from literature have been corrected to the cosmology adopted in this work. The $L_{\rm 2500\AA}$ collected from literatures were estimated from the $M_{\rm 1450\AA}$ by assuming a power-law slope of $\alpha_\nu=-0.5$, while the $L_{\rm 2500\AA}$ for $z>6.5$ quasars are directly measured from quasar spectra. This plot indicates that there is no redshift evolution of the $\rm \alpha_{ox} - L_{\rm 2500\AA}$ relation.
\label{fig_ox_l2500}}
\end{figure}

The $\alpha_{\rm ox}$ measurement traces the relative importance of the disk emission versus corona emission and is an important parameter for investigating the accretion physics of luminous quasars \citep[e.g.][]{Brandt15}. Previous studies have shown that there is a tight correlation between $\alpha_{\rm ox}$  and $L_{\rm 2500 \AA}$ \citep[e.g.][]{Just07, Lusso16}. \cite{Nanni17} and \cite{Vito19} recently used more measurements of high-redshift quasars
and showed that the $\alpha_{\rm ox}-L_{\rm 2500\AA}$ relation does not depend on redshift. 

\begin{figure}[tbh]
\centering
\includegraphics[width=0.48\textwidth]{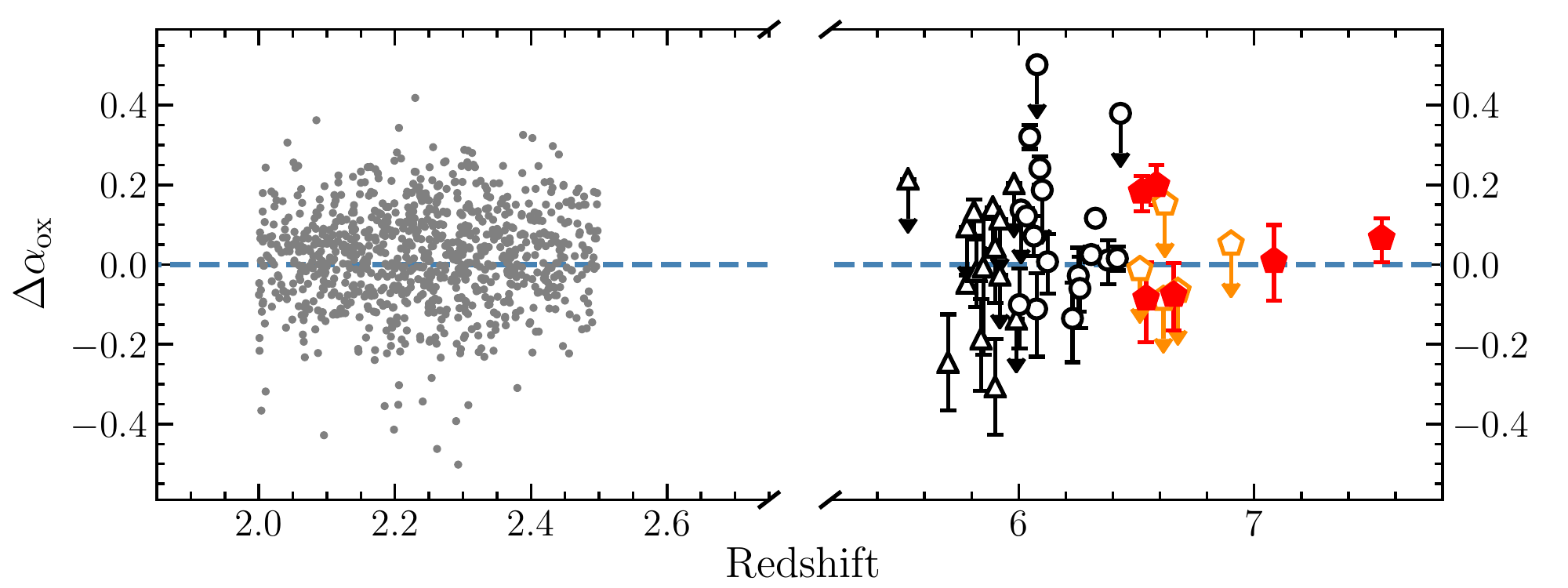}
\caption{The $\rm \Delta \alpha_{ox}$ versus redshift plot. All symbols have the same meaning as Figure \ref{fig_ox_l2500}.
\label{fig_delta_ox}}
\end{figure}

We further investigate the $\alpha_{\rm ox}-L_{\rm 2500\AA}$ relationship of high-redshift quasars in three redshift bins with $5.5<z<6.0$ quasars from \cite{Nanni17}, $6.0<z<6.5$ quasars from \cite{Vito19}, and $z>6.5$ quasars from our analysis, which are shown in Figure \ref{fig_ox_l2500}. In this Figure, we also plot $\alpha_{\rm ox}-L_{\rm 2500\AA}$ 
measured by fitting lower redshift quasars \citep{Just07,Martocchia17,Timlin20} as well as from $z\sim6$ quasars \citep{Nanni17}. Our analysis agrees with previous work  \citep[e.g.][]{Nanni17, Banados18a, Vito19} showing a tight  $\alpha_{\rm ox}-L_{\rm 2500\AA}$ relation for quasars at different redshifts. In Figure \ref{fig_delta_ox}, we show the relation between $\rm \Delta \alpha_{ox}$, the difference between the measured $\rm \alpha_{ox}$ and the value expected from the \cite{Timlin20} $\alpha_{\rm ox}-L_{\rm 2500\AA}$ relation, and quasar redshift. At all redshifts, the $\rm \Delta \alpha_{ox}$ is distributed around zero, indicating that there is no redshift evolution of the $\alpha_{\rm ox}-L_{\rm 2500\AA}$ relationship up to $z\sim7$. 
Since $L_{\rm 2500\AA}$ is proportional to $L_{\rm bol}$ (because the $L_{\rm bol}$ was estimated from $L_{\rm 3000\AA}$), and $\alpha_{\rm ox}$ is a relation between $L_{\rm 2500\AA}$ and \xray\ luminosity, the lack of redshift evolution of the $\alpha_{\rm ox}-L_{\rm 2500\AA}$ relation is fully consistent with the discussion in \S \ref{subsec_kbol} about the redshift independent relationship between $k_{\rm bol}$ and $L_{\rm bol}$. 

\begin{figure}[tbh]
\centering
\includegraphics[width=0.49\textwidth]{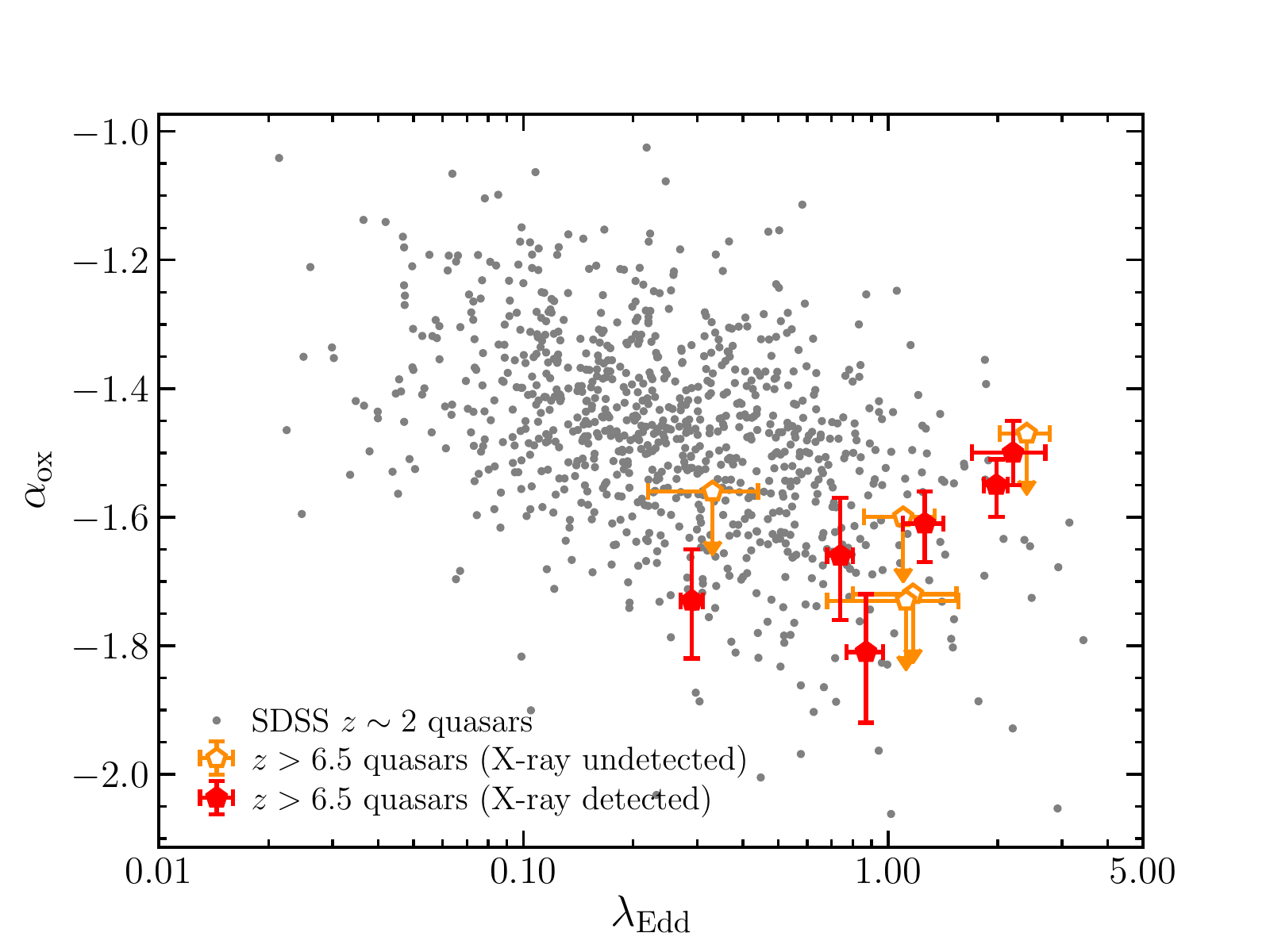}
\caption{
The correlation between $\alpha_{\rm ox}$ and Eddington ratio, $\lambda_{\rm Edd}$. The Spearman test gives $\rho=-0.43$ and $p=2.3\times10^{-41}$, suggesting a moderate correlation between $\alpha_{\rm ox}$ and $\lambda_{\rm Edd}$.  All symbols have the same meaning as Figure \ref{fig_ox_l2500}.
\label{fig_ox_redd}}
\end{figure}

The Eddington ratio $\lambda_{\rm Edd}$ is the relative accretion rate of the SMBH. 
\cite{Shemmer08} and \cite{Lusso10} found that a weak correlation exists between $\alpha_{\rm ox}$ and $\lambda_{\rm Edd}$ from the analyses of $\sim$30 and $\sim$150 quasars at lower redshifts, respectively. In order to test whether  $\alpha_{\rm ox}$ depends on the $\lambda_{\rm Edd}$ at high redshift, we also correlate $\alpha_{\rm ox}$ with $\lambda_{\rm Edd}$ for the 897 SDSS quasars from \cite{Timlin20} and the $z>6.5$ quasars in Figure \ref{fig_ox_redd}. Although the relation shows large scatter, a Spearman test gives a correlation coefficient of $\rho=-0.43$ and a chance probability of $p=2.3\times10^{-41}$, suggesting a moderate $\alpha_{\rm ox}$ - $\lambda_{\rm Edd}$ relation and that the $\alpha_{\rm ox}$ steepens with increasing $\lambda_{\rm Edd}$. Our analysis shows a stronger $\alpha_{\rm ox}$ - $\lambda_{\rm Edd}$ relation compared with that in  \cite{Shemmer08} and \cite{Lusso10}, which could be a natural result of the improved statistics arising from a much larger quasar sample. Nevertheless, the dispersion of this relation is still significant due to the large uncertainty on the individual $\lambda_{\rm Edd}$ measurements, which has a systematic uncertainty up to $\sim0.55$ dex from the $M_{\rm BH}$ estimate\citep{Vestergaard09}. On the other hand, we need to keep in mind that $\lambda_{\rm Edd}$ correlates with $L_{\rm 2500\AA}$, thereby the $\alpha_{\rm ox}$ - $\lambda_{\rm Edd}$ relation could also be a consequence of the inherent dependence of $\lambda_{\rm Edd}$ and $L_{\rm 2500\AA}$ as suggested by \cite{Shemmer08}. 
 
\begin{figure}[tbh]
\centering
\includegraphics[width=0.49\textwidth]{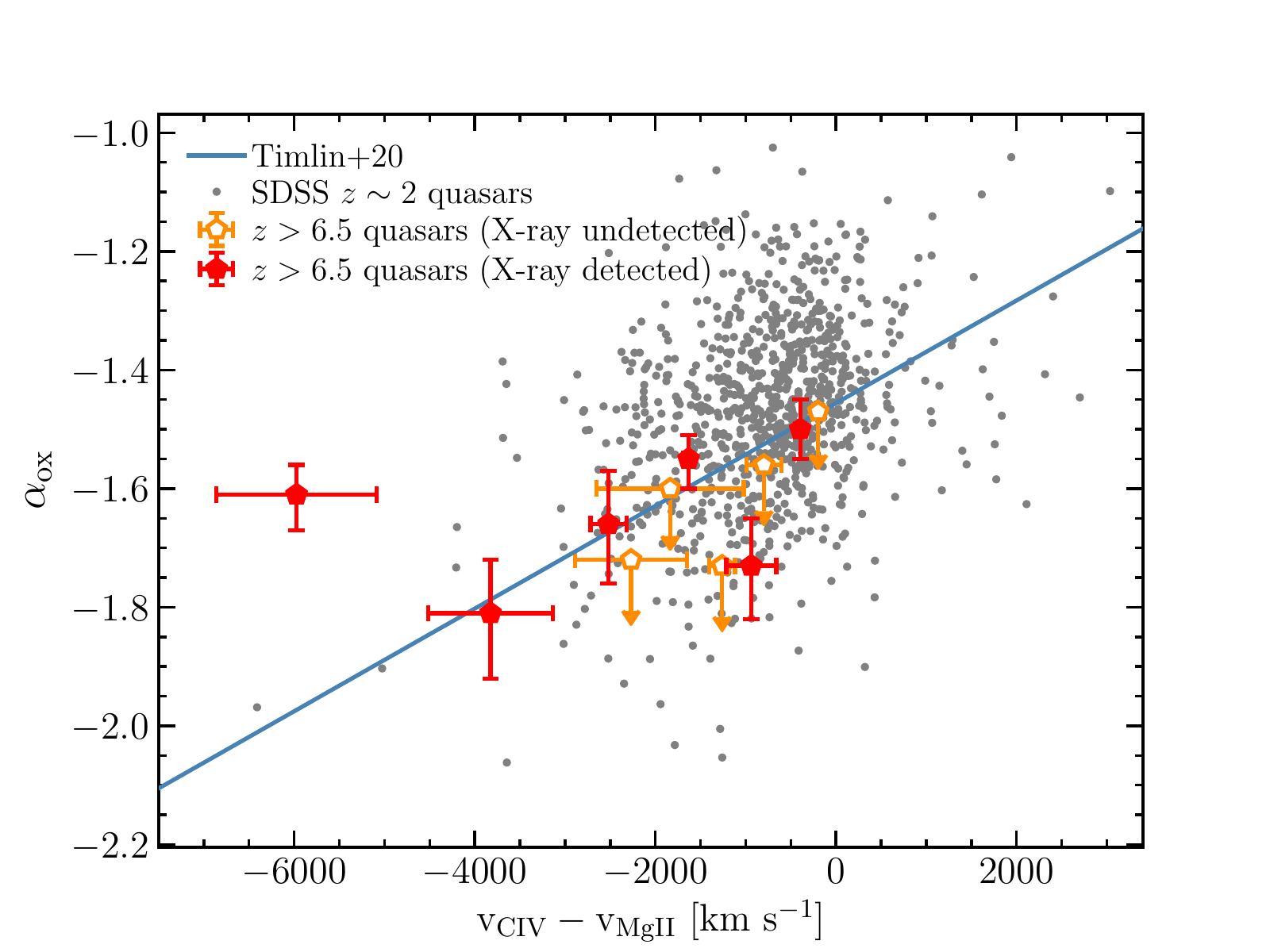}
\caption{
The correlation between $\alpha_{\rm ox}$ and \Civ\ line blueshifts. The Spearman test on all quasars shown in this figure gives $\rho=0.32$ and $p=1.2\times10^{-23}$, suggesting a moderate relation between these two quantities.  All symbols have the same meaning as Figure \ref{fig_ox_l2500}.
\label{fig_ox_blueshift}}
\end{figure}

In addition to the broad band SED shape, the relative importance of \xray\ and UV emission can also affect the radiation driven wind from the accretion disk, where the \xray\ photons can strip the gas of electrons and thereby reduce the line driving, while the UV photons accelerate the wind due to radiation line pressure \citep[the so-called disk+wind model; e.g.][]{Proga00, Richards11}. Therefore, the relatively soft spectrum (smaller $\alpha_{\rm ox}$) would drive a strong wind \citep[e.g.][]{Kruczek11}. As such,  $\alpha_{\rm ox}$ is an important parameter for understanding the radiation driven wind. It is commonly suggested that the blueshift of high-ionization broad emission lines, like \Civ,  is a marker for radiation driven winds launched from the accretion disk \citep[e.g.][]{Gaskell82,Richards11}. Thus, one would expect the $\alpha_{\rm ox}$ to be correlated with \Civ\ line blueshift. 
Indeed, a moderate correlation between  $\alpha_{\rm ox}$ and \Civ\ line blueshift have been found at low redshifts \citep[e.g.][]{Richards11, Timlin20} which supports the paradigm discussed above.  
On the other hand, recent studies  found that the \Civ\ line blueshift of the most distant quasars is about a factor of $\sim$ 2.5 larger than that of lower redshift quasars \citep[e.g.][]{Mazzucchelli17,Meyer19,Schindler20}. Investigations of whether the most distant quasars follow the $\alpha_{\rm ox}$ and \Civ\ line blueshift relation found in lower redshift quasars will give us more insights on whether the radiation driven wind in quasars evolves with redshift. In Figure \ref{fig_ox_blueshift}, we show the relation between $\alpha_{\rm ox}$ and \Civ\ line blueshift for both $z>6.5$ quasars and SDSS lower redshift quasars. The blueshifts were derived from the redshifts of \Mgii\ and \Civ\ lines as listed in Table \ref{tab_nir}. In this Figure, the $z>6.5$ quasars show higher \Civ\ line blueshifts than most of SDSS quasars, but they still follow the blue line derived by \cite{Timlin20} based solely on SDSS $z\sim2$ redshift quasars. Although J1342+0928, the most distant quasar in our sample, is far from the relation found by \cite{Timlin20}, such outliers in SDSS quasars with smaller blueshift velocities are also seen in this plot. A larger sample of quasars at $z\sim7$ with both \xray\ and NIR observations are needed to shed more light on this question. 

\subsection{X-ray vs Infrared Luminosity}

\begin{figure}
\centering
\includegraphics[width=0.49\textwidth]{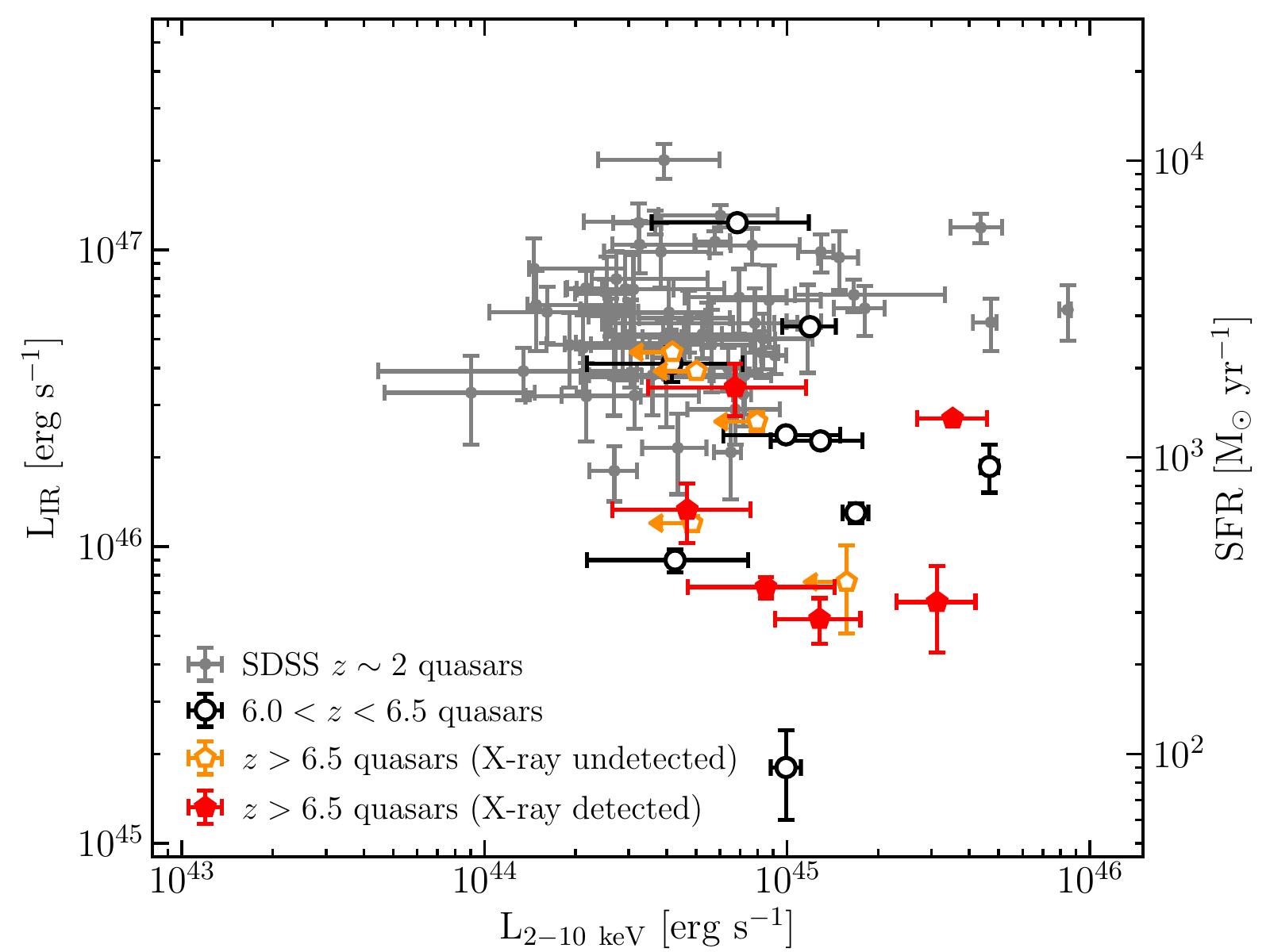}
\caption{ Infrared luminosity, $L_{\rm IR}$, versus \xray\ luminosity, $L_{\rm 2-10keV}$. The Spearman test gives $\rho=-0.19$ and p=0.10, suggesting no correlation between the star formation rate (as indicated by $L_{\rm IR}$) and AGN activity (as indicated by $L_{\rm 2-10keV}$), different from that found in low redshift AGNs \citep[e.g.][]{Alexander05,Netzer09,Xu15}. All symbols have the same meaning as Figure \ref{fig_ox_l2500}.
\label{fig_lx_irx}}
\end{figure}

The observed relations between the masses of SMBHs and the masses of the bulges in their host galaxy suggest a connection between SMBHs and their host galaxies \citep[see][for a review]{Kormendy13}. The underlying relation between the average host star formation and AGN luminosity found in low redshift high luminosity AGNs \citep[e.g.][]{Alexander05,Netzer09,Xu15} leads to a relationship between bulge and SMBH growth rates. Recent work by \cite{Rosario12} finds that the relation between star formation and AGN activity in luminous AGNs weakens or disappears at high redshifts ($z>1$), suggesting an evolutionary relation between SMBH and host galaxy growth rates at high redshifts. In order to investigate whether the quasar \xray\ properties (i.e. \xray\ luminosity) correlate with quasar host galaxy properties (i.e. $L_{\rm IR}$ or SFR) in the earliest epochs, we plot the $L_{\rm IR}$ and $L_{\rm 2-10keV}$ of all objects as described in \S\,\ref{sec_measure} in Figure \ref{fig_lx_irx}. 
From Figure \ref{fig_lx_irx},  there is no correlation ($\rho=-0.19$, p=0.10) between $L_{\rm IR}$ (or SFR) and $L_{\rm 2-10keV}$ for both SDSS $z\sim2$ quasars and high-$z$ quasars, different from that in lower redshift ($z<1$) AGNs \citep[e.g.][]{Netzer09,Xu15}. The lack of a correlation between $L_{\rm IR}$ and $L_{\rm 2-10keV}$ of these luminous quasars is also consistent with the absence of a correlation between $L_{\rm IR}$ and $L_{\rm bol}$ of $z\gtrsim6$ quasars \citep{Venemans18} and the high mass ratio between SMBHs and their host galaxies \citep[e.g.][]{Decarli18, Wang19a}, indicating that SMBHs of the most luminous quasars in the early epochs do not co-evolve with their host galaxies, at least not following the same relation found in low redshift galaxies \citep[e.g.][]{Alexander05,Netzer09,Kormendy13}. We emphasize that our quasar sample only represents the UV brightest quasar population at high redshift and the conclusion can only apply to these most luminous objects.

Moreover, Figure \ref{fig_lx_irx} shows that the $L_{\rm IR}$ of the most luminous (e.g. $L_{\rm 2-10keV}\gtrsim10^{45}~{\rm erg~s^{-1}}$) $z>6.5$ quasars is even fainter than that of \xray\ fainter ones (e.g. $L_{\rm 2-10keV}<10^{45}~{\rm erg~s^{-1}}$) and most SDSS quasars. However, as we mentioned in  \S\,\ref{sec_measure}, the {\it Herschel} detected SDSS quasars only represent the $\sim$15\% infrared bright quasars limited by the depth of {\it Herschel} observations, thus some of the {\it Herschel} undetected SDSS quasars could have similar infrared-to-\xray\ luminosity ratios with $z>6.5$ quasars. Nevertheless, there is no $z>6.5$ quasar having $L_{\rm IR}$ close to $\rm 10^{47} erg~s^{-1}$ and several \xray\ bright high-$z$ quasars with  $L_{\rm IR}<10^{46} {\rm erg~s^{-1}}$ suggests that these powerful AGN with strong disk driven wind, as indicated by the high Eddington ratio (Figure \ref{fig_ox_redd}) and large \Civ\ blueshift (Figure \ref{fig_ox_blueshift}), could suppress star formation activities in their host galaxies. However, our current sample is too small to obtain definitive conclusions and a systematic survey of the \xray\ and far-infrared properties of a larger $z>6.5$ quasar sample would be critical to test this scenario. 

\section{Average X-ray Properties of $z\sim7$ Quasars} \label{sec_stack}
\begin{figure}[tbh]
\centering
\includegraphics[width=0.48\textwidth]{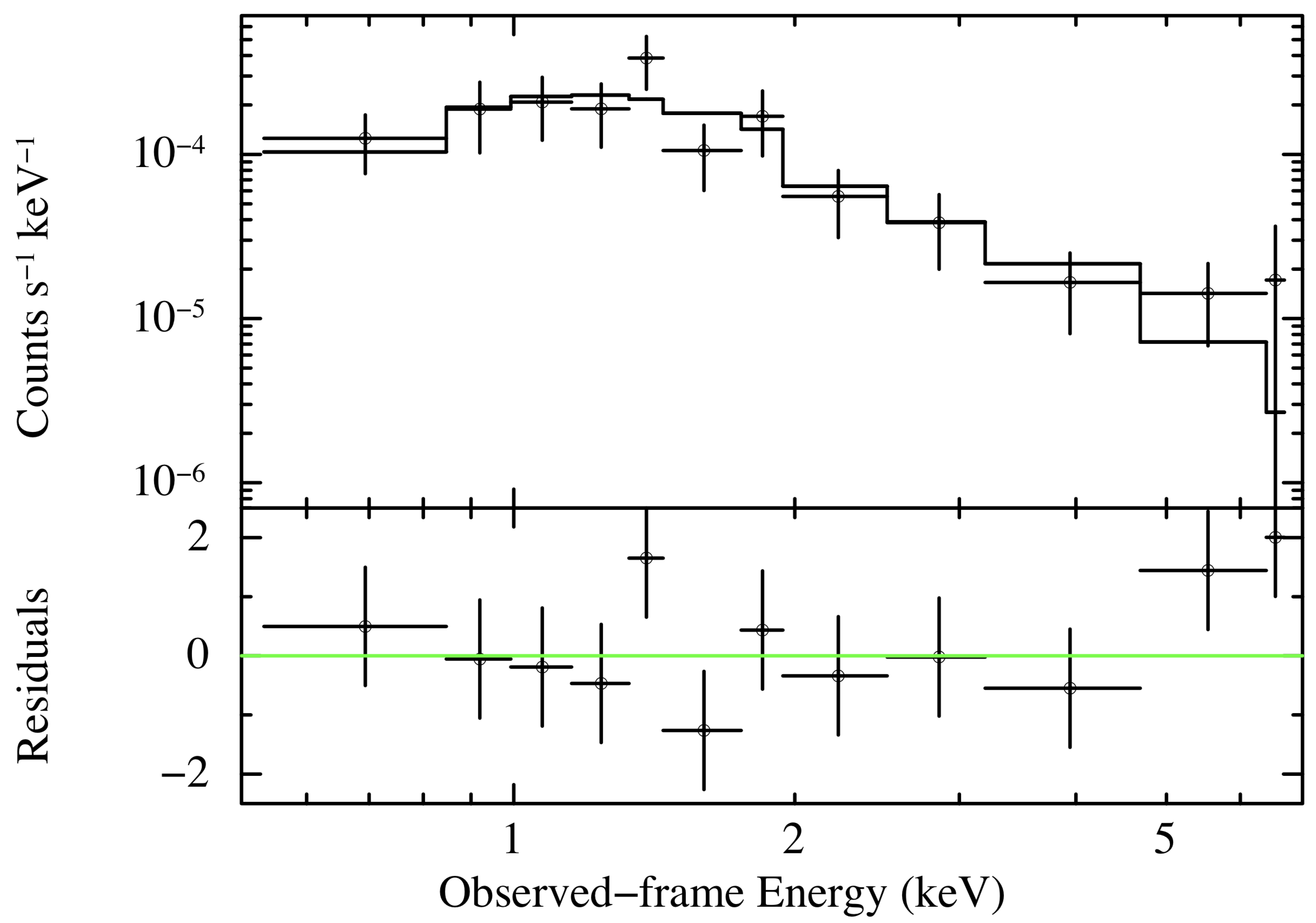}
\caption{Stacked \xray\ spectrum of six \xray\ detected $z>6.5$ quasars. The mean redshift of these quasars is $z=6.822$. The black solid line is the best-fit power-law model with a photon index of $\Gamma=2.11^{+0.27}_{-0.26}$, consistent with $\Gamma=2.32^{+0.31}_{-0.30}$, derived from the joint spectral fitting. The bottom panel shows the residuals (data$-$model).
\label{fig_stack}}
\end{figure}

Measuring the \xray\ spectral properties for individual quasars requires a significant number of detected \xray\ counts. In this work, we do not attempt to fit the individual quasar \xray\ spectrum due to the limitation of small number of photons detected. Instead, we measure the average hard \xray\ photon index of these $z>6.5$ quasars using two different methods. First, we perform a joint spectral analysis of the six $z>6.5$ quasars that are well detected in the \xray\ (see Table \ref{tab_cxo}). The average redshift of these six quasars is $z=6.822$ and the total net counts in the 0.5--7 keV band is $\sim$64. 
We jointly fit these quasar spectra with a power-law model and associate a value of redshift and Galactic absorption to each source using {\tt XSPEC}. From the joint fit, we derive a photon index $\Gamma = 2.32^{+0.31}_{-0.30}$. We use the Cash statistic and report the uncertainties at the 68\% confidence level. As a further test, we stack the spectra of these six detected $z>6.5$ quasars. The stacked spectrum is shown in Figure \ref{fig_stack}. We use {\tt XSPEC} to fit this stacked spectrum with a power-law by fixing the Galactic absorption component to the mean $N_{\rm H}$ and the redshift to $z=6.822$. The derived photon index from the stacked spectrum fitting is $\Gamma=2.11^{+0.27}_{-0.26}$, consistent with the photon index obtained by the joint spectral fitting. 

\begin{figure*}[tbh]
\centering
\includegraphics[width=0.48\textwidth]{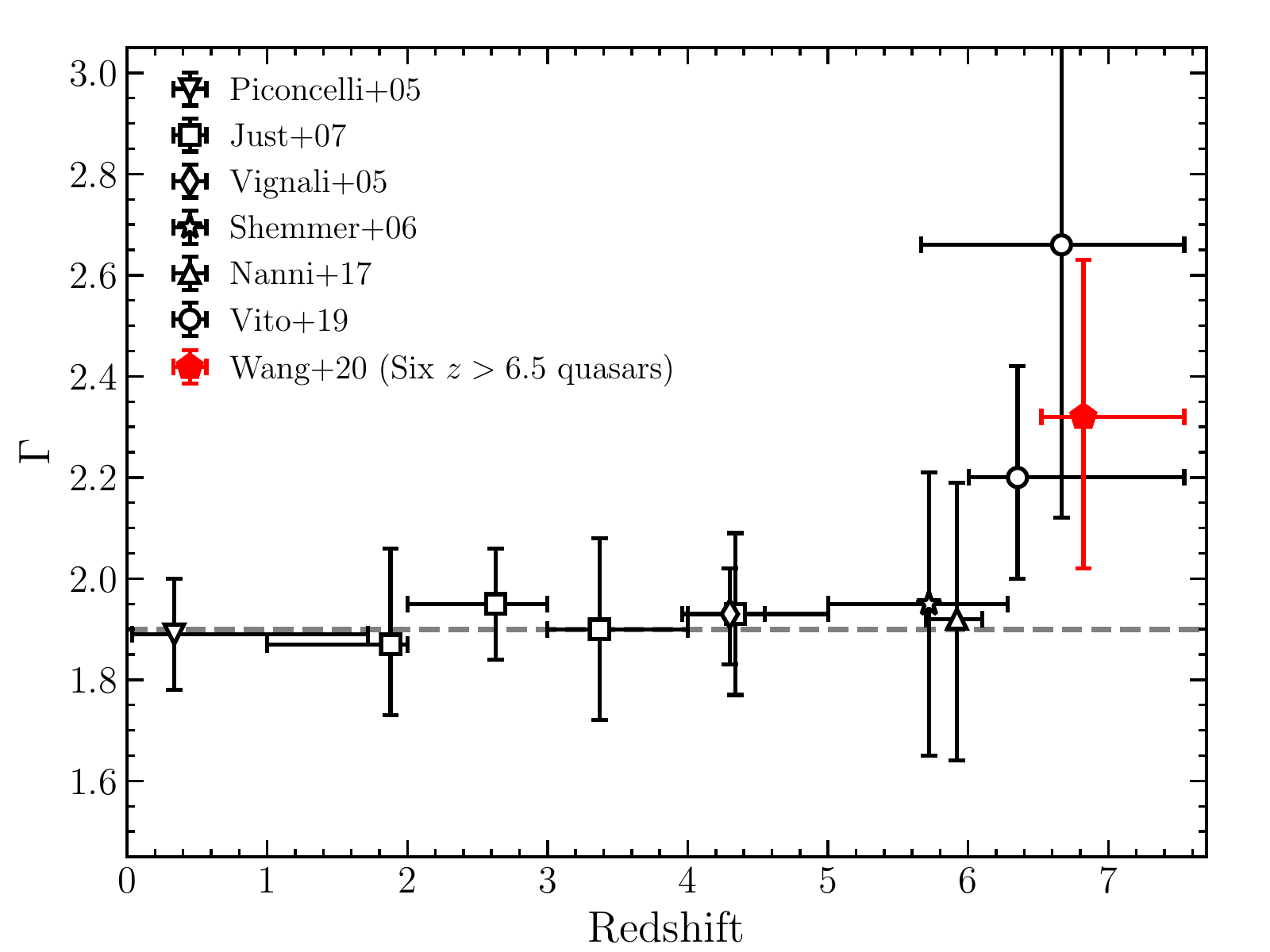}
\includegraphics[width=0.48\textwidth]{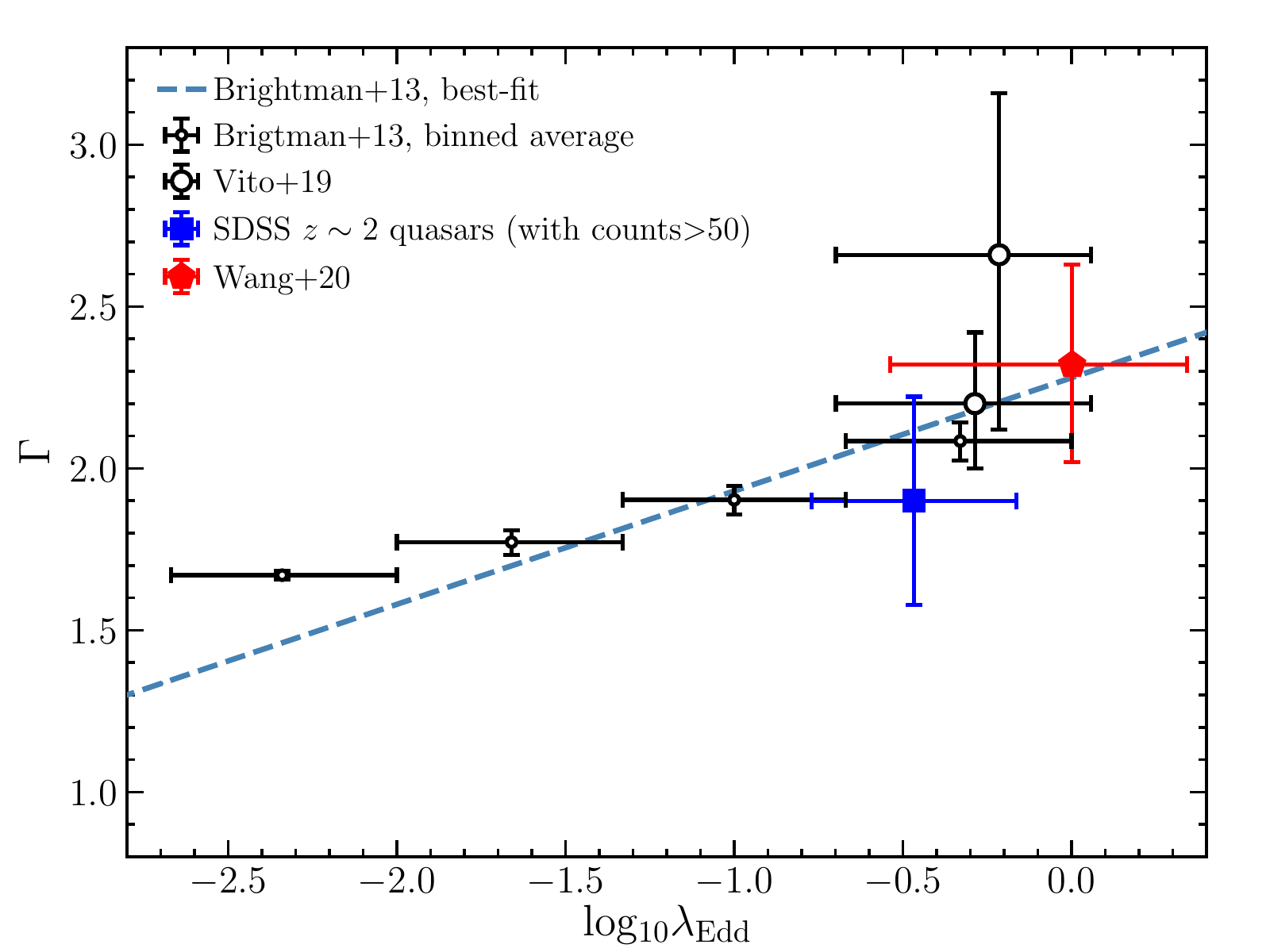}
\caption{{\it Left}: Photon index, $\Gamma$, as a function of redshift. Our joint spectra fitting result ($\Gamma=2.32^{+0.31}_{-0.30}$) at $z=6.822$ is highlighted with a red pentagon. The gray dashed line denotes $\rm \Gamma=1.9$. In all cases, the assumed model is a power-law and errors are reported at the 68\% confidence level.
{\it Right}: Photon index, $\Gamma$, as a function of Eddington ratio, $\lambda_{\rm Edd}$. Our new measurement based on the joint fitting of six \xray\ detected $z>6.5$ quasars is highlighted as a red pentagon. The two measurements by \cite{Vito19} are shown as big open circles. The small open circles denote binned averages measurements of a sample of $z\lesssim2$ quasars with both \xray\ observations and BH mass estimates and the dashed line represents the best-fit model \citep{Brightman13}.  
\label{fig_gamma}}
\end{figure*}

\cite{Vito19} jointly analyzed three $z>6.5$ quasars ($\sim$23 net counts in total) and found  $\Gamma=2.66^{+0.54}_{-0.50}$. The average $\Gamma$ measured by \cite{Vito19} is slightly steeper than (although with large uncertainties) the average $\Gamma$ found at lower redshifts, which is $\Gamma\sim1.9$ \citep{Piconcelli05,Vignali05,Shemmer06,Just07,Nanni17}. In the left panel of Figure \ref{fig_gamma}, we show the average $\Gamma$ measured from joint spectral fitting of quasars at different cosmic epochs. 
Our newly measured $\Gamma$ is slightly steeper than that for lower redshift quasars,  consistent with the value measured by \cite{Vito19} but with smaller uncertainties.

Since the $\Gamma$ value of a quasar's hard \xray\ spectrum correlates with the $\lambda_{\rm Edd}$ value as suggested by numerous works \citep[e.g.][]{Porquet04, Shemmer08, Brightman13}, it is necessary to check whether the steeper hard \xray\ spectral slope at $z>6.5$ is due to quasars with high Eddington ratios. Since not all quasars studied in these joint spectral analyses \citep[e.g.][]{Porquet04, Shemmer08, Brightman13} have Eddington ratio measurements, we can not compare the Eddington ratios of these quasars used for joint spectral analyses directly. Instead, we compare the average $\Gamma$ and $\lambda_{\rm Edd}$ of our $z>6.5$ quasars with the relation found by \cite{Brightman13} from a well studied AGN sample at $0.5\lesssim z \lesssim 2.0$ in the Cosmic Evolution Survey (COSMOS) and Extended Chandra Deep Field South (E-CDF-S) field. In the right panel of Figure \ref{fig_gamma} we show the measurements and relation from \cite{Brightman13} as well as our measurement at $z>6.5$. In this Figure, we also show the average $\Gamma$ and $\lambda_{\rm Edd}$ of the $z\gtrsim6$ quasars from \cite{Vito19} and  SDSS $z\sim2$ quasars. Note that the average $\Gamma$ of $z\sim2$ SDSS quasars is directly measured from the \xray\ spectral fitting by \cite{Timlin20} and only includes $\sim100$ quasars with $>50$ net counts selected from our $z\sim2$ comparison quasar sample (see \S \ref{sec_correlation}). This plot indicates that the steeper hard \xray\ slope of the $z>6.5$ quasars from our analysis and the previous study by \cite{Vito19}  are fully consistent with the $\Gamma$ and $\lambda_{\rm Edd}$ relation found in lower-$z$ quasars, suggesting that the steeper $\Gamma$ of $z>6.5$ quasars is mainly driven by their higher Eddington ratios rather than by their higher redshifts.

\section{Summary} \label{sec_summary}
In this paper, we present new \cxo\ observations of five quasars at $z>6.5$. By combining them with archival \cxo\ observations of an additional six $z>6.5$ quasars, we perform a systematic analysis of the \xray\ properties of these reionization-era quasars. Six of these eleven $z>6.5$ quasars are well detected with a luminosity range of $L_{\rm 2-10 keV}\sim (4.7-35.3)\times10^{44}~{\rm erg~s^{-1}}$.
In addition, we analyze the infrared spectroscopic observations of these \cxo\ observed $z>6.5$ quasars and derive the bolometric luminosities, BH masses, Eddington ratios, and broad emission line blueshifts for all quasars. The bolometric luminosities of these sources span a range of $L_{\rm bol}\sim (0.5-3.4)\times10^{47}~{\rm erg~s^{-1}}$, occupying the bright end of the quasar population. Their masses and Eddington ratios are in the range  $(0.2-3.0)\times10^{9}~M_\odot$ and $\sim0.3-2.4$, respectively. We also measure the infrared luminosity ($L_{\rm IR}$) and star formation rate (SFR) of the quasar host galaxies yielding $L_{\rm IR}$ in the range of $(0.5-4.5) \rm \times10^{46}~ erg~s^{-1}$ and SFR in the range of $\sim 200-2000~M_\odot ~{\rm yr^{-1}}$, respectively. Moreover, we perform a joint spectral analyses of all \xray\ detected quasars and measure the average \xray\ spectral properties of these $z>6.5$ quasars. Our findings from this unique sample of $z>6.5$ quasar with both \xray\ and near-infrared spectroscopic observations, and based on a comparison quasar sample at $z\sim2$, are as follows:

\begin{itemize}

\item The \xray\ bolometric luminosity correction ($k_{\rm bol}=L_{\rm bol}/L_{\rm 2-10 keV}$) of $z>6.5$ quasars increases with bolometric luminosity and the optical/UV to \xray\ flux ratio, $\alpha_{\rm ox}$, strongly correlates with quasar luminosity at rest-frame 2500 \AA, $L_{\rm 2500 \AA}$, following the same trend found in lower redshift quasars.

\item A moderate correlation between $\alpha_{\rm ox}$ and Eddington ratio, $\lambda_{\rm Edd}$, exists. This correlation is weaker than the $\alpha_{\rm ox}$-$L_{\rm 2500 \AA}$ relation, which could either be a consequence of the inherent dependence of  $\lambda_{\rm Edd}$ and $L_{\rm 2500 \AA}$ or result from the large uncertainty introduced by the $\lambda_{\rm Edd}$ measurement. 

\item The $L_{\rm IR}$ and SFR do not correlate with the $L_{\rm 2-10 keV}$ in these luminous distant quasars, suggesting that the ratio of the SMBH growth rate and their host galaxy growth rate in these early luminous quasars are different from that of local galaxies.

\item There is a moderate correlation between $\alpha_{\rm ox}$ and \Civ\ line blueshift. In the disc+wind model picture \citep[e.g.][]{Gaskell82, Richards11}, the \Civ\ line blueshift increases as the relative importance of corona \xray\ emission and accretion disk emission decreases, consistent with the observed correlation.

\item The average photon index, $\Gamma$,  of hard \xray\ spectra of $z>6.5$ quasars is found to be $\Gamma=2.32^{+0.31}_{-0.30}$, steeper than that of lower redshift quasars. By comparing our measurement with the $\Gamma$-$\lambda_{\rm Edd}$ relation found in lower redshift quasars \citep[e.g.][]{Brightman13}, we conclude that the steeper $\Gamma$ of $z > 6.5$ quasars is mainly driven by their higher Eddington ratios rather than by their higher redshifts. 

\end{itemize}

In the near future, a larger sample of $z>6.5$ quasars with both \xray, NIR, and sub-millimeter observations, as well as a well matched (in terms of both quasar luminosity and observational depth) quasar sample at lower redshifts is critical for investigating whether the earliest SMBHs are fed by different accretion physics (especially the \xray\ luminosity and $\Gamma$) and arise in distinct galactic environments (i.e. star formation rate) relative to their lower redshift counterparts.

\acknowledgments
We thank M. Brightman for kindly providing their photon index and Eddington ratio measurements, thank V. D'Odorico for approving the use of their unpublished ESO archival data., and  thank F. Vito for providing his collection of photon index measurements at different redshifts. We thank Richard Green and Jianwei Lyu for useful discussion. 
We thank the anonymous referee for reading the paper carefully and providing useful comments. 
Support for this work was provided by NASA through the NASA Hubble Fellowship grant \#HST-HF2-51448.001-A awarded by the Space Telescope Science Institute, which is operated by the Association of Universities for Research in Astronomy, Incorporated, under NASA contract NAS5-26555. X. Fan and J. Yang acknowledge support from the US NSF Grant AST-1515115 and NASA ADAP Grant NNX17AF28G. Support for this work was provided by the National Aeronautics and Space Administration through Chandra Award Number GO8-19079X issued by the Chandra X-ray Center, which is operated by the Smithsonian Astrophysical Observatory for and on behalf of the National Aeronautics Space Administration under contract NAS8-03060.

The scientific results reported in this article are based to a significant degree on observations made by the Chandra X-ray Observatory and data obtained from the Chandra Data Archive. This research has made use of software provided by the Chandra X-ray Center (CXC) in the application packages CIAO, ChIPS, and Sherpa. The scientific results reported in this article are based in part on observations obtained at the international Gemini Observatory, acquired through the Gemini Observatory Archive at NOIRLab, which is managed by the Association of Universities for Research in Astronomy (AURA) under a cooperative agreement with the National Science Foundation. On behalf of the Gemini Observatory partnership: the National Science Foundation (United States), National Research Council (Canada), Agencia Nacional de Investigaci\'{o}n y Desarrollo (Chile), Ministerio de Ciencia, Tecnolog\'{i}a e Innovaci\'{o}n (Argentina), Minist\'{e}rio da Ci\^{e}ncia, Tecnologia, Inova\c{c}\~{o}es e Comunica\c{c}\~{o}es (Brazil), and Korea Astronomy and Space Science Institute (Republic of Korea). The scientific results reported in this article are based in part  on observations collected at the European Southern Observatory under ESO programmes 087.A-0890(A), 097.B-1070(A), 098.B-0537(A), and 0100.A-0625(A).
This paper makes use of the following ALMA data: ADS/JAO.ALMA\#2018.1.01188.S. ALMA is a partnership of ESO (representing its member states), NSF (USA) and NINS (Japan), together with NRC (Canada), MOST and ASIAA (Taiwan), and KASI (Republic of Korea), in cooperation with the Republic of Chile. The Joint ALMA Observatory is operated by ESO, AUI/NRAO and NAOJ.
The National Radio Astronomy Observatory is a facility of the National Science Foundation operated under cooperative agreement by Associated Universities, Inc.

\vspace{5mm}
\facilities{ALMA, CXO, Gemini(GNIRS), Magellan(FIRE), VLT(X-SHOOTER)}
\software{ ChiPS \citep{chips}, CIAO \citep{Fruscione06}, PypeIt \citep{pypeit1,pypeit2}, Sherpa \citep{sherpa}, XSPEC \citep{xspec}}

\end{document}